\documentclass{jpsj3}

\title{Disorder Effects on Competition between Antiferromagnetism and Superconductivity in 
Cuprate Superconductors through the Enhancement in Charge Susceptibility}

\author{Hiromasa \textsc{Tamaki}
\footnote
{
Present Address: 
Advanced Technology Research Laboratories, Panasonic Corporation, 
3-4 Hikaridai, Seika, Kyoto 619-0237, Japan
}
 and Kazumasa \textsc{Miyake}}

\inst{Division of Materials Physics, 
Department of Materials Engineering Science, Graduate School of Engineering Science, 
Osaka University, Toyonaka, Osaka 560-8531, Japan}

\recdate
{
\today
}

\abst{
The coexistence state of antiferromagnetism (AF) and superconductivity (SC) has been observed in 
five-layered cuprates. However, this coexistence state disappears, and the AF phase and SC phase lose 
contact in the doping phase diagram toward double- and single-layered cuprates.
We investigate the mechanism of the disappearance of the coexistence of AF and SC in disordered 
cuprate superconductors in order to understand these doping phase diagrams. 
In single- and double-layered cuprates, electrons on the CuO$_2$ plane experience the disorder effect 
through inhomogeneity in the charge reservoir layer.  These impurity potentials can be effectively enhanced toward 
the underdoped region by the effect of many-body corrections that involve an increase in charge 
susceptibility.  As a result, strong disorder effects are expected particularly in the competing regions of 
AF and SC, where the coexistence phase of AF and SC is extremely suppressed.
We show the validity of this suppression mechanism by considering the Aslamazov-Larkin-type vertex correction to the 
effective impurity potential in the effective mean-field phase diagram.
}
\kword{cuprate superconductor, competition of antiferromagnetism and superconductivity, 
disorder, charge compressibility, Aslamazov-Larkin term, Maki-Thompson term}

\begin{document}
\maketitle

\section{Introduction}
%%high-tc cuprate
%%coexistence of AF and SC
The general structure of doping phase diagrams in the underdoped region is one of the 
critical issues with 
high-temperature cuprate superconductors. In single-layered cuprates such as La$_{2-x}$Sr$_x$CuO$_4$ 
(LSCO), which have been studied extensively since its discovery, it has been known that the 
antiferromagnetic (AF) phase and superconducting (SC) phase do not coexist and are 
separately located in the hole-doping phase diagram while sandwiching a spin-glass 
phase. \cite{Keimer1992}
However, it has recently been discovered by nuclear magnetic (quadrupole) resonance experiments 
that a five-layered cuprate superconductor shows the coexistence of the AF and 
SC states in the underdoped region. \cite{Mukuda2006a,Mukuda2006b,Mukuda2008,Mukuda2009}
The discovery of such coexistence indicates that a pure CuO$_2$ plane shows the coexistence of the AF 
and SC phases, and that the single-layered cuprates such as LSCO lose their coexistent phase 
because of disorder from the charge reservoir layer adjacent to the CuO$_2$ plane.
%theory of coexistence
The existence of the coexistent phase has been substantiated by numerous theoretical approaches 
on the basis of the Hubbard model \cite{Reiss2007,Jarrell2001,Aichhorn2006,Aichhorn2007,
Senechal2005,Kobayashi2009,Lichtenstein2000,Capone2006,Kancharla2008} 
and $t$-$J$ model. \cite{Inaba1996,Yamase2004,Himeda1999,Pathak2009a} 
%theory of disappearance of coexistence?
However, there has been no established theory for describing the disappearance of the coexistence due 
to inhomogeneity or some other factors on a unified picture.

%out-of-plane disorder
The existence of disorder effects from the charge reservoir layer, the so-called out-of-plane 
disorder 
effects, has been reported both theoretically and experimentally.
The impurity potential from the Sr site in LSCO is estimated by the local spin-density 
approximation +U (LSDA+U) method. \cite{Anisimov1992} LSDA+U studies have shown that the in-gap 
band with an oxygen-like character shifts further into the gap by $\sim 1{\rm eV}$ through 
a distortion of 
the CuO$_2$ plane. In comparison with the hopping energy estimated as $t_{\rm dp}\sim 1{\rm eV}$, 
this impurity potential through the lattice distortion induced by the out-of-plane disorder is not strong enough to reach the unitarity limit 
but is also ineligible for the strong disturbance of the electronic state. This situation is different from 
an in-plane disorder such as the Zn substitution of the Cu site where Zn itself becomes 
a stronger scatterer.
%experiments
In the case of the in-plane disorder, the  suppression of the SC transition temperature 
$T_{\rm c}$ in 
the underdoped region is sharper than that in the overdoped region.\cite{Harashina1993}
The suppression of the SC transition temperature $T_{\rm c}$ due to an out-of-plane substitution 
has been observed in single-layered cuprates. \cite{Fujita2005,Sugimoto2006,Okada2008,Hobou2009}
Its suppression of $T_{\rm c}$ occurs with a smaller residual resistivity in the case of the 
out-of-plane disorder than in the case of the in-plane disorder. \cite{Fujita2005} 
The pseudogap temperature $T^*$ 
is robust against the out-of-plane disorder in contrast to the $T_{\rm c}$ being sensitive to 
the out-of-plane disorder. \cite{Okada2008}

%%schematic phase diagram
%%%%%%%%%%%%%%%%%%%%%%%%%%%%%%%%%%%%%%%%%%%%%%%%%%%%%%%%%%%%%%%%%%%%%%%%%%%%%%
\begin{figure}
 \begin{center}
  \includegraphics[scale=0.2]{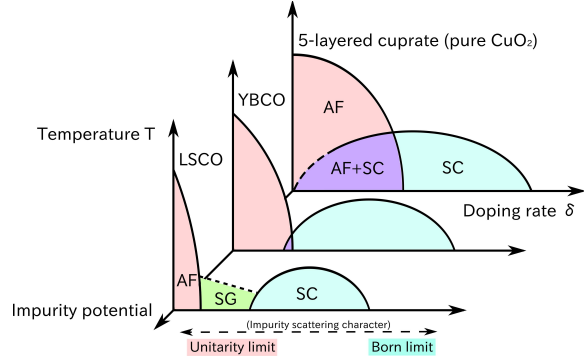}
 \end{center}
\caption{Schematic doping-temperature phase diagram for different disorder strengths we propose.  
Antiferromagnetism (AF), superconductivity (SC), their coexistence (AF+SC), and spin-glass (SG) phase 
are drawn. Here, the stripe phase at $1/8$-filling and incommensurate SDW are not drawn for simplicity.  
We discuss the difference in phase diagrams among the cuprates from the viewpoint of the strength of 
the disorder potential on the CuO$_2$ plane.}
 \label{fig-schematic}
\end{figure}
%%%%%%%%%%%%%%%%%%%%%%%%%%%%%%%%%%%%%%%%%%%%%%%%%%%%%%%%%%%%%%%%%%%%%%%%%%%%%%%
In the double-layered cuprate YBa$_2$Cu$_3$O$_{6+x}$ (YBCO), whose CuO$_2$ plane is less disordered 
than the LSCO but not so clean as the five-layered cuprate, the coexistence of a spin density wave 
and superconductivity has recently been observed by neutron scattering. \cite{Haug2010}
Figure \ref{fig-schematic} shows a schematic phase diagram of the five-layered, double-layered, 
and single-layered cuprates by taking the disorder strength of the CuO$_2$ plane into account.
We illustrate in their phase diagrams that the AF state, SC state, and their coexistence 
state (AF+SC) are suppressed around the competing regions of the AF, SC, and 
AF+SC states accompanied by the effect of impurity potential. In LSCO, not only does 
the AF+SC phase vanish but also the AF phase and SC phase lose contact with each other 
because of the stronger impurity potential resulting from the substitution of La by Sr.
However, the difference between the phase diagrams shown in Fig. \ref{fig-schematic} does 
not seem to have been clarified yet on the basis of the microscopic theory.  In this paper, we propose a possible 
mechanism to describe the difference in doping phase diagrams among clean and disordered 
cuprate superconductors.

This paper is organized as follows: 
In \S2, we introduce the concept of renormalized impurity potential and consider its behavior 
in cuprate. In \S3, we discuss the role of AL-type correction in charge susceptibility 
and impurity scattering. Then, on the basis of the microscopic theory starting from the 
Hubbard model, we analyze the effects of the Aslamazov-Larkin (AL) term on charge susceptibility 
(\S4), and on the phase diagram of disordered cuprates (\S5 and \S6).
%dimensionality effect?

\section{Relationship of Fermi Liquid Corrections between Charge Susceptibility and 
Impurity Potential}
%AG formula
A nonmagnetic impurity potential generally suppresses antiferromagnetism and unconventional 
superconductivity. The impurity-concentration dependence of the SC transition temperature $T_{\rm c}$ 
is well described by the well-established Abrikosov-Gor'kov formula in many
superconductors. \cite{Abrikosov1961}

In this section, we introduce the concept of the effective impurity potential renormalized 
by electron-electron correlation effects. This concept paves the way for discussing disorder effects 
such as the suppression of AF and SC in cuprates, which are typical strongly correlated electron systems.

%fermi liquid relation using mass renormalization and Landau parameter
%impurity potential renormalization 
%%%%%%%%%%%%%%%%%%%%%%%%%%%%%%%%%%%%%%%%%%%%%%%%%%%%%%%%%%%%%%%%%%%%%%%%%%%%%%%
\begin{figure}
 \begin{center}
  \includegraphics[scale=0.4]{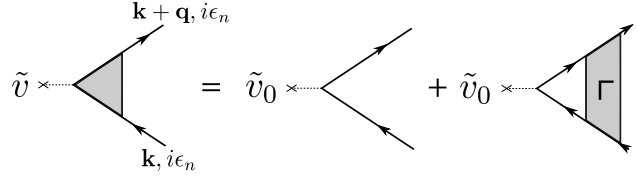}
 \end{center}
 \caption{Diagrammatic representation of relationship between the effective impurity 
potential $\tilde{v}$ and the bare potential $v_0$ through the correction from the vertex 
function $\Gamma$.}
\label{fig-imp_pot-vertex}
\end{figure}
%%%%%%%%%%%%%%%%%%%%%%%%%%%%%%%%%%%%%%%%%%%%%%%%%%%%%%%%%%%%%%%%%%%%%%%%%%%%%%%
The renormalized impurity potential $\tilde{v}$ is defined by considering the vertex correction as
\begin{eqnarray}
 \tilde{v}({\bf k},{\bf k}+{\bf q},i\epsilon_n)=
v_0({\bf q})\Lambda({\bf k},{\bf q},i\epsilon_n),
\end{eqnarray} 
with
\begin{eqnarray}
 \Lambda({\bf k},{\bf q},i\epsilon_n)=1+T\sum_{{\bf k}',n'}G({\bf k}',i\epsilon_n')
G({\bf k}'+{\bf q},i\epsilon_n')\Gamma({\bf k}',i\epsilon_n',{\bf k},i\epsilon_n,;{\bf q},0), 
\label{eq:lambda}
\end{eqnarray}
where $v_0$ and $\Lambda$ are the bare impurity potential and its renormalization factor, 
respectively. $\Gamma$ denotes a four-point vertex function. Figure \ref{fig-imp_pot-vertex} shows 
the diagrammatic representation of the renormalized impurity potential $\tilde{v}$. 
The renormalization factor of the impurity scattering for the quasi-particle at the Fermi surface 
is given 
in the forward scattering limit ${\bf q}=0$ \cite{Betbeder-Matibet1966}:
\begin{eqnarray}
\lim_{{\bf q}\rightarrow 0}\Lambda({\bf k}_{\rm F},{\bf q},0)=\frac{1}{z}\frac{1}{1+F_0^s},
\label{eq:renorm-imp}
\end{eqnarray}
where 
$1/z\equiv 1-\partial\Sigma^{\rm R}({\bf k}_{\rm F},\epsilon)/\partial\epsilon|_{\epsilon=0}$ is 
the mass renormalization factor and $F_0^s$ is the isotopic and spin-symmetric component of the 
Landau parameters.

A similar enhancement factor can be seen in the charge susceptibility given by the 
linear response theory as
\begin{eqnarray}
 \chi_{\rm c}({\bf q},i\omega_m)=-\int_0^{\beta} d\tau\left<n_{\bf q}(\tau)n_{-{\bf q}}(0)\right>
e^{i\omega_m\tau},
\end{eqnarray}
where $n_{\bf q}(\tau)=\sum_{{\bf k},\sigma}c^{\dagger}_{{\bf k+q}\sigma}
(\tau)c_{{\bf k}\sigma}(\tau)$ is the charge density operator. $c^{\dagger}_{{\bf k}\sigma}$ and 
$c_{{\bf k}\sigma}$ are the creation and annihilation operators for the particle with the momentum 
${\bf k}$ and spin $\sigma=\uparrow,\downarrow$, respectively.
We can define the charge compressibility $\kappa\equiv \partial n/\partial \mu$, which is the 
uniform limit of the static charge susceptibility:
\begin{eqnarray}
\kappa= \lim_{{\bf q}\rightarrow 0}\chi_{\rm c}({\bf q},\omega=0).\label{eq:kappa-rel}
\end{eqnarray}
In a noninteracting system, the charge compressibility reaches $\kappa_0\simeq2N_{\rm F}$ 
where $N_{\rm F}$ is the density of states at a Fermi level.
The relation of the charge compressibility $\kappa$ with $\kappa_0$ can be derived from Fermi 
liquid theory:~\cite{Landau1956}
\begin{eqnarray}
\frac{\kappa}{\kappa_0}
=\frac{m^*}{m}\frac{1}{1+F_0^s},\label{eq:kappa-landau}
\end{eqnarray}
where $m^*/m$ is the mass enhancement factor, which is generally different from $1/z$ by the 
$k$-mass appearing from the ${\bf k}$-dependence of the self-energy of electrons.  
The $k$-mass cancels the mass renormalization factor $1/z$. 
In La$_{1-x}$Sr$_x$MnO$_3$, for example, the effective mass $m^*$ remains unrenormalized, while the 
Drude weight $D$ decreases toward the metal-insulator transition. At the transition, 
the Drude weight $D=\pi n^*e^2/m^*$ becomes zero owing to the disappearance of the effective carrier 
density $n^*\equiv n(1+F_{1}^{\rm s}/3)$.~\cite{Miyake2001} 
However, if the enhancement of $\Lambda$ and $\kappa$ originate from the anomaly in the 
Landau parameter $F_0^s\rightarrow -1$, 
we can identify the origin of the enhancement in the charge compressibility as that in the 
effective impurity potential in the 
forward scattering limit.

In the presence of quantum critical fluctuations, vertex corrections affect the enhancement in 
the effective impurity potential and residual 
resistivity. \cite{Miyake2002a,Miyake2002b,Maebashi2002}
As discussed above, the enhancement (or suppression) factor of the impurity potential is related to 
that of the charge compressibility.
For these reasons, we can expect a strong disorder effect in the material in which a large 
enhancement in the charge compressibility is observed.

%Report of enhanced charge fluctuations

The charge compressibility, or uniform charge susceptibility, has been studied both theoretically 
and experimentally in cuprate superconductors.
%\subsection{Monte Carlo method}
Pioneering theoretical works on the charge compressibility around the metal-insulator transition 
have been performed with the use of the quantum Monte Carlo 
method. \cite{Otuka1990,Furukawa1991,Furukawa1992,Furukawa1993}.
They have shown in the two-dimensional Hubbard model that the charge compressibility diverges as 
the doping rate to the Mott insulating phase decreases as $\delta\equiv 1-n\rightarrow 0$.  
The charge compressibility behaves as $\kappa\propto 1/\delta$ around the Mott insulating phase. 
This result implies that the transition from a metal to a Mott insulator induced by controlling the 
filling is of the second order.  Furukawa {\it et al.} have also concluded that the divergence 
$\kappa$ involves the behavior of the Landau parameter, 
$F_0^{\rm s}\rightarrow -1$ as $\delta\rightarrow 0$. Then, they have speculated that the 
effective mass $m^*$ diverges.
The $1/\delta$ dependence in $\kappa$ has also been obtained from the path-integral renormalization 
group (PIRG) method.~\cite{Watanabe2004a} Since the quantum Monte Carlo and PIRG methods 
are applicable to systems with up to about 100 sites, the appearance of the singular 
charge susceptibility is not limited in small size systems. 

The experimental study of the charge compressibility has been conducted  by photoemission 
spectroscopy. \cite{Ino1997,Harima2001,Harima2003,Yagi2006,Ikeda2010}
The theoretical prediction of the chemical potential shift $\Delta\mu\propto \delta^2$, which 
yields $\kappa\equiv-\partial \delta/\partial \Delta\mu \propto 1/\delta$, is satisfied in LSCO, 
Bi-2212, and YBCO within the range of experimental error.

Although half-filled cuprates are charge transfer insulators with a different nature from the 
Mott-Hubbard insulator appearing in the Hubbard model, the enhancement in the charge susceptibility 
is observed in some cuprates.
Their results suggest that a weak impurity scattering potential should be enhanced to the extent 
of the unitarity limit in the underdoped region of LSCO, Bi-2212, and YBCO, where an enhanced charge 
compressibility is observed.

\section{Aslamazov-Larkin Term in Charge Susceptibility and Impurity Potential}\label{sec:3}
%AL fluctuations
%%%%%%%%%%%%%%%%%%%%%%%%%%%%%%%%%%%%%%%%%%%%%%%%%%%%%%%%%%%%%%%%%%%%%%%%%%%%%%%
\begin{figure}
 \begin{center}
  \includegraphics[scale=0.8]{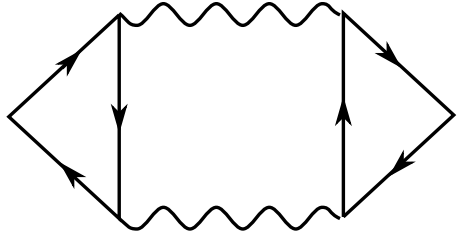}
 \end{center}
 \caption{AL term in the charge susceptibility. Solid lines and wavy lines represent the 
single-particle Green function and spin fluctuation propagator, respectively.}
\label{fig-AL-term}
\end{figure}
%%%%%%%%%%%%%%%%%%%%%%%%%%%%%%%%%%%%%%%%%%%%%%%%%%%%%%%%%%%%%%%%%%%%%%%%%%%%%%%
In this section, we introduce an approach to describing the singular behavior of the charge 
susceptibility on the basis of the mode-mode coupling theory.

The divergence of the charge susceptibility in the Hubbard model is known to involve the divergence 
of the AF spin correlation length from quantum Monte Carlo 
calculations. \cite{Furukawa1991,Furukawa1992}  The singular behavior of the charge susceptibility 
has also been obtained in a one-dimensional Hubbard model by the cellular dynamical mean field 
theory (DMFT), but this singularity cannot be obtained by the single-site DMFT. \cite{Capone2004} 
This suggests the importance of an off-site correlation for the 
appearance of the singular charge susceptibility.
There is little doubt that the appearance of the singular charge susceptibility in the Hubbard 
model is closely related to the Mott transition. However, we cannot exclude the possibility 
of the development of AF correlation as 
the origin of the singular charge susceptibility with this background.

It is desirable to understand the mechanism of the enhancement in the charge susceptibility from 
the structure of the single-particle excitation spectrum in the Green function $G$ and the 
four-point vertex function $\Gamma$ in eq. (\ref{eq:lambda}).
From this viewpoint, it has been known that the contribution of the vertex correction 
shown in Fig. \ref{fig-AL-term}, which is 
the so-called AL term, gives a singularity to the charge susceptibility around the AF critical 
point. \cite{Miyake1994}
The AL-type contribution to the static charge susceptibility constructed by the 
two-spin-fluctuation propagator $\chi_{\rm s}({\bf q},i\omega_m)$ is given by
\begin{eqnarray}
 \Delta\chi_{\rm c,AL}({\bf q})=3T\sum_{{\bf q}',m}\left[\gamma_3({\bf q},{\bf q}';
i\omega_m)\right]^2\chi_{\rm s}({\bf q}',i\omega_m)\chi_{\rm s}({\bf q}'+{\bf q},i\omega_m),
\label{eq:chicAL}
\end{eqnarray}
where the factor $3$ corresponds to the number of spin directions. The effect of the AL-type 
diagram has been discussed in light of the problem of the electric conductivity anomaly in 
the presence of strong SC fluctuations. \cite{Aslamazov1968}
$\gamma_3$ denotes the coupling strength among one charge fluctuation field $n(q)$ and two 
spin fluctuation fields $S_i(q)$ ($i=x, y, z$), and consist of three Green functions $G$'s for 
electrons:
\begin{eqnarray}
 \gamma_3({\bf q},{\bf q}';i\omega_m)=\lambda^2 T\sum_{{\bf k},n}G({\bf k},i\epsilon_n)
G({\bf k}+{\bf q},i\epsilon_n)G({\bf k}+{\bf q}',i\epsilon_n+i\omega_m),
\end{eqnarray}
where $\lambda$ is the cubic coupling constant in the vertex with the structure of 
$\sum_{k,q}S_i(q)c^{\dagger}(k)c(k+q)$.

Here, we evaluate the effect of critical spin fluctuations that are singular near the 
incommensurate AF wave vector ${\bf Q}^*_i$ in the form
\begin{eqnarray}
\chi_{\rm s}({\bf q},i\omega_m)=
\sum_i\frac{N_{\rm F}}{\eta_{\rm s}+A({\bf q}-{\bf Q}^*_i)^2+C|\omega_m|},\label{eq:chis}
\end{eqnarray}
where $\eta_{\rm s}$ is the distance from the magnetic critical point.
The spin susceptibility in doped cuprate shows its peaks at four incommensurate ordering vectors 
${\bf Q}^*_i=\pi/a(1,1\pm\epsilon)$, $\pi/a(1\pm\epsilon,1)$.

Since the dominant contribution to $\kappa_{\rm AL}({\bf q})$ arises from the summation at 
$\omega_m=0$ and around ${\bf q}'\sim{\bf Q}^*$ in the case of $\eta_{\rm s}\ll 2C\pi T$, 
the contribution of the coherent part denoted by eq. (\ref{eq:chis}) is estimated as
\begin{eqnarray}
 \Delta\chi_{\rm c,AL}({\bf q})=3T\sum_{i,j}\gamma_3({\bf q},{\bf Q}^*_i;0)
\gamma_3({\bf q},{\bf Q}^*_j;0)\sum_{\bf q'}\frac{N_{\rm F}}{\eta_{\rm s}+A({\bf q}'-{\bf Q}^*_i)^2}
\frac{N_{\rm F}}{\eta_{\rm s}+A({\bf q}+{\bf q}'-{\bf Q}^*_j)^2}.
\end{eqnarray}
We can see that $\Delta\chi_{\rm c,AL}({\bf q})$ shows its peak at 
${\bf q}={\bf Q}^*_i-{\bf Q}^*_j$.  The maximum is attained at ${\bf q}=0$:
\begin{eqnarray}
\Delta\kappa_{\rm AL}&=&\Delta\chi_{\rm c,AL}(0)\nonumber\\
&=&3\times4\left[\gamma_3(0,{\bf Q}^*;0)\right]^2T\sum_{{\bf q}'}\frac{N_{\rm F}^2}
{(\eta_{\rm s}+Aq'^2)^2}\nonumber\\
&\simeq&
\begin{cases}
\frac{3\left[\gamma_3(0,{\bf Q}^*;0)\right]^2}{\pi A}\cdot\frac{N_{\rm F}^2T}{\eta_{\rm s}}
&({\rm 2D,\ }T\gg \eta_{\rm s}/2C\pi)\\

\frac{3\left[\gamma_3(0,{\bf Q}^*;0)\right]^2}{2\pi A^{3/2}}\cdot\frac{N_{\rm F}^2T}
{\sqrt{\eta_{\rm s}}}
&({\rm 3D,\ }T\gg \eta_{\rm s}/2C\pi)
\end{cases}.\label{eq:AL-finiteT}
\end{eqnarray}
In the two-dimensional square lattice with only the nearest-neighbor hopping, 
$\gamma_3(0,{\bf Q}^*;0)$ takes a nonzero value except for the half-filling (Appendix A).

At low temperatures, $\eta_{\rm s}\gg 2C\pi T$, $\Delta\kappa_{\rm AL}$ shows a less singular 
behavior because the finite-${\omega_m}$ contribution is not negligible. At zero temperature $T=0$, 
the frequency summation in eq. (\ref{eq:chicAL}) can be approximated by the integration with respect 
to $\omega$. As in the case of $\eta_{\rm s}\gg 2C\pi T$, the maximum is attained at ${\bf q}=0$:
\begin{eqnarray}
 \Delta\kappa_{\rm AL}&\simeq&3\times4\left[\gamma_3(0,{\bf Q}^*;0)\right]^2 \cdot 2\sum_{{\bf q}'}
\int_0^{\infty}\frac{d\omega}{2\pi}\frac{N_{\rm F}^2}{(\eta_{\rm s}+Aq'^2+C\omega)^2}\nonumber\\
&=&24\left[\gamma_3(0,{\bf Q}^*;0)\right]^2\frac{N_{\rm F}^2}{C}\sum_{{\bf q}'}\frac{1}
{\eta_{\rm s}+Aq'^2}\nonumber\\
&\simeq&
\begin{cases}
\frac{6\left[\gamma_3(0,{\bf Q}^*;0)\right]^2N_{\rm F}^2}{\pi AC}\log\frac{Aq_{\rm c}}{\eta_{\rm s}}
&({\rm 2D,\ }T=0)\\
\frac{12\left[\gamma_3(0,{\bf Q}^*;0)\right]^2N_{\rm F}^2q_{\rm c}}{\pi^2 AC}
\left(1-\frac{q_c}{2\pi^2}\sqrt{\frac{\eta_{\rm s}}{A}}\right)
&({\rm 3D,\ }T=0)
\end{cases},\label{eq:AL-zeroT}
\end{eqnarray}
where $q_{\rm c}\sim 1/a$ is the short-wavelength cutoff and of the order of the inverse of the 
lattice constant $a$. The singularity in $\Delta\kappa_{\rm AL}$ at quantum critical points is 
weakened in two dimensions or suppressed to the cusp singularity in three dimensions.
The charge susceptibility shows the power-law singularity near the magnetic instability 
$\eta_{\rm s}\sim 0$ at finite temperatures, where
the critical exponent of $\Delta\kappa_{\rm AL}$ are the same as that of the SDW 
susceptibility $\chi_{\rm s}({\bf Q}^*,0)$ in two dimensions, and half of that of 
$\chi_{\rm s}({\bf Q}^*,0)$ in three dimensions.
However, this critical exponent is not always reliable because it is easily changed by other 
processes such as the mode-mode coupling effect among charge-density channels.

Other peaks in $\Delta\chi_{\rm c,AL}({\bf q})$ appear at ${\bf q}={\bf Q}^*_i-{\bf Q}^*_j\neq 0$.
These peaks are generally smaller 
than $\Delta\chi_{\rm c,AL}(0)$ because $\gamma_3$ is smaller for such mode coupling processes.  
The case of ${\bf Q}^*_i=-{\bf Q}^*_j$ corresponds to a stripe charge fluctuation. The stripe charge 
fluctuations are known to be widely observed in cuprate superconductors. Moreover, a stripe ordered 
phase is stabilized instead of the SC phase at a doping rate $\delta=1/8$. We note that 
$\Delta\chi_{\rm c,AL}({\bf q})$ shows the emergence of stripe fluctuations, but unfortunately, is 
not accountable for the stabilization of the stripe ground state with completely suppressed 
superconductivity only at $\delta=1/8$ in cuprates.

%damping rate and resistivity
%%%%%%%%%%%%%%%%%%%%%%%%%%%%%%%%%%%%%%%%%%%%%%%%%%%%%%%%%%%%%%%%%%%%%%%%%%%%%%%
\begin{figure}
 \begin{center}
  \includegraphics[scale=1]{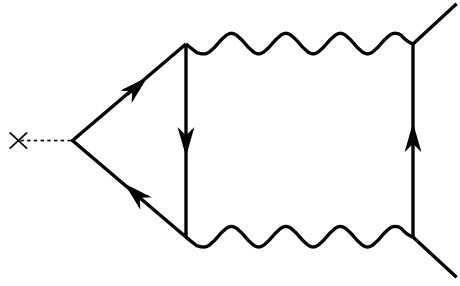}
 \end{center}
 \caption{AL term in impurity scattering process.}
\label{fig-imp-AL}
\end{figure}
%%%%%%%%%%%%%%%%%%%%%%%%%%%%%%%%%%%%%%%%%%%%%%%%%%%%%%%%%%%%%%%%%%%%%%%%%%%%%%%
Here, we consider the effect of the AL process on the impurity scattering problem. By taking into 
account the AL-type correction to the impurity scattering process, we can obtain a singularly 
strong disorder effect around AF critical points. Such a correction must be expressed in 
diagrammatic form as shown in Fig. \ref{fig-imp-AL}.

When the charge susceptibility is nearly critical at ${\bf q}=0$, the renormalized impurity potential 
$\tilde{v}$ at the Fermi level can be expanded around the forward scattering as 
\begin{eqnarray}
 \tilde{v}({\bf q})
\sim \frac{1}{\eta_{\rm c}+A|{\bf q}|^2}v_0({\bf q}),
\end{eqnarray}
where $\eta_{\rm c}$, the distance from the critical point with respect to a uniform charge density 
fluctuation, corresponds to the square inverse of the charge density correlation length 
$\eta_{\rm c}\propto 1/\xi_{\rm c}^2$. This kind of vertex correction leads to a long-ranged 
impurity potential even if the bare potential $v_0$ is short-ranged. The range of the impurity 
potential is given by $\xi_{\rm c}$, which diverges at the critical point.

We discuss the scattering rate of quasi-particles and the residual resistivity due to renormalized 
impurity scattering under the condition of a spherical Fermi surface.
The scattering rate $1/\tau$, which causes the suppression of the superconductivity and 
antiferromagnetism, relates to the imaginary part of the self-energy. It is attained for the 
impurity concentration $c_{\rm imp}$ within the Born approximation as 
\begin{eqnarray}
 \frac{1}{\tau_{\bf k}}&\simeq&2\pi c_{\rm imp}\sum_{{\bf k}'}\left|\tilde{v}({\bf k}-{\bf k}')\right|^2
 z\delta(\xi^*({\bf k}'))\nonumber\\
&=&2\pi N_{\rm F}c_{\rm imp}\left<\frac{v_0^2}{\left(\eta_{\rm c}+2Ak_{\rm F}^2
(1-\cos \theta)\right)^2}\right>_{\rm FS},\label{eq:tau}
\end{eqnarray}
where $\xi^*({\bf k})$ is the quasi-particle dispersion and $\theta$ is the angle between 
${\bf k}$ and {\bf k}'. $\left<\cdots\right>_{\rm FS}$ denotes the average on the Fermi surface. 
Here, note that the Born approximation is not applicable in the strong impurity scattering region
around the criticality $\eta_{c}\sim0$. In such cases, eq. (\ref{eq:tau}) should be replaced with 
$1/\tau_{\bf k}=2 c_{\rm imp}{\rm Im}\tilde{t}({\bf k},{\bf k})$ with the use of the $t$-matrix 
$\tilde{t}({\bf k},{\bf k'})\equiv \tilde{v}({\bf k}-{\bf k}')
+\sum_{\bf k''}\tilde{v}({\bf k}-{\bf k}'')G_{\rm R}({\bf k}'',0)\tilde{t}({\bf k}'',{\bf k'})$.
If we ignore the momentum dependence in $\tilde{v}({\bf k}-{\bf k}')$, the $t$-matrix is 
approximated as $\tilde{t}^{-1}=\tilde{v}^{-1}-i\pi N_{\rm F}$.
By this reason, the scattering amplitude saturates to $1/\tau\sim 2c _{\rm imp}/\pi N_{\rm F}$ 
toward the unitarity limit. 
However, we can examine the feature of the enhancement even if the Born
approximation is adopted. By taking the average in eq. (\ref{eq:tau}), the behavior of 
the scattering rate near the criticality $\eta_{\rm c}\sim 0$ is
\begin{eqnarray}
 \frac{1}{\tau_{{\bf k}_{\rm F}}}\propto
\begin{cases}
1/\eta_{\rm c}^{3/2} & {\rm (2D)}\\
1/\eta_{\rm c} & {\rm (3D)}
\end{cases}.
\end{eqnarray}  

The electric resistivity shows a different behavior from the scattering rate.
The residual resistivity $\rho_0$ due to the impurity scattering is obtained as 
\begin{eqnarray}
 \rho_0&\propto&\frac{1}{\tau_{\rm tr}}\nonumber\\
&=&2\pi N_{\rm F}c_{\rm imp}\left<\frac{v_0^2(1-\cos\theta)}{\left(\eta_{\rm c}+2Ak_{\rm F}^2
(1-\cos \theta)\right)^2}\right>_{\rm FS}.
\end{eqnarray}
The contribution of the forward scattering is subtracted in $\rho_0$ so that the lifetime in 
transport phenomena, $\tau_{\rm tr}$, is different from $\tau_{\bf k}$ in general. We can easily take 
this average.  As a result, we obtain 
\begin{eqnarray}
 \rho_0\propto 
\begin{cases}
1/\sqrt{\eta_{\rm c}} & {\rm (2D)} \\
\log(1/\eta_{\rm c}) & {\rm (3D)}
\end{cases}.
\end{eqnarray}
The residual resistivity also shows a divergence at critical point $\eta_{\rm c}=0$.
However, its power of divergence is weak compared with that of the scattering rate 
$1/\tau_{k_{\rm F}}$.  
Thus, even if the enhanced impurity potential, due to critical charge fluctuations, 
destroys the long-range order, the residual resistivity does not show a visible enhancement in 
certain cases. Therefore, even though the residual resistivity $\rho_0$ is insensitive to 
the impurity potential in contrast with the quasi-particle lifetime $\tau_{\bf k}$, 
the SC transition temperature $T_{\rm c}$ can vanish at a smaller residual resistivity 
in the underdoped region than in the overdoped region.

\section{Aslamazov-Larkin-type correction to the charge susceptibility in the Hubbard model}
In this section, we analyze the role of the AL-type vertex correction in the Hubbard model with the 
use of the spin-fluctuation propagator at the level of random phase approximation (RPA). 
The consideration of the spin susceptibility within the RPA level is consistent with the analysis 
of the mean-field phase diagram for the antiferromagnetism.  For comparison, we also study 
the effect of the Maki-Thompson (MT) term consisting of aq diagram simpler than the AL term.

%Hubbard model
The two-dimensional Hubbard model in a square lattice is described by
\begin{eqnarray}
 H=\sum_{{\bf k},\sigma}\epsilon({\bf k})c^{\dagger}_{{\bf k}\sigma}c_{{\bf k}\sigma}+
U\sum_{\bf q,k,k'}c^{\dagger}_{{\bf k+q}\uparrow}c_{{\bf k}\uparrow}c^{\dagger}_{{\bf k'}\downarrow}
c_{{\bf k'+q}\downarrow}\, .\label{eq:Hubbard}
\end{eqnarray}
The dispersion in cuprates is well fitted by 
$\epsilon({\bf k})=-2t(\cos k_x+\cos k_y)+4t'\cos k_x\cos k_y$ with the hopping integral between 
the nearest-neighbor sites $t$ and the next-nearest neighbor sites $t'$.  In this study, we choose 
$t'=0$ for simplicity and owing to the limitation of RPA analysis (see \S 6).
In this section, we perform a diagrammatic analysis by putting the Green function for noninteracting 
electrons $G^{(0)}_{\sigma}({\bf k},i\epsilon_n)=\left[i\epsilon_n-\epsilon({\bf k})+\mu\right]^{-1}$ 
into all $G_{\sigma}({\bf k},i\epsilon_n)$. This means that no self-energy correction is considered 
in the Green function $G$.  

%RPA-based vertex corrections

%%%%%%%%%%%%%%%%%%%%%%%%%%%%%%%%%%%%%%%%%%%%%%%%%%%%%%%%%%%%%%%%%%%%%%%%%%%%%%%
\begin{figure}
 \begin{center}
  \includegraphics[scale=0.7]{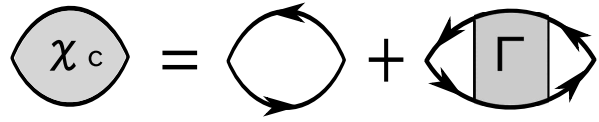}
 \end{center}
 \caption{Diagrammatic representation of relationship between charge susceptibility and 
vertex function.}
\label{fig-chic-vertex}
\end{figure}
%%%%%%%%%%%%%%%%%%%%%%%%%%%%%%%%%%%%%%%%%%%%%%%%%%%%%%%%%%%%%%%%%%%%%%%%%%%%%%%
The charge susceptibility is expressed in terms of the vertex function $\Gamma$ and the Green function $G$ as shown in Fig. \ref{fig-chic-vertex}, and its analytic expression is given as 
\begin{eqnarray}
 \chi_{\rm c}(q)=2\chi_{0}(q)+T^2\sum_{k,k',\sigma,\sigma'}G_{\sigma}(k)G_{\sigma}(k+q)
\Gamma_{\sigma\sigma'}(k,k';q)G_{\sigma'}(k')G_{\sigma'}(k'+q),
\end{eqnarray}
where we have introduced the abbreviations $k=({\bf k},i\epsilon_n)$ and $q=({\bf q},i\omega_m)$ with 
the Matsubara frequencies, $\epsilon_{n}=(2n+1)\pi T$ for electrons, and  $\omega_m=2\pi mT$ 
for bosonic excitations.  We define the irreducible particle-hole bubble $\chi_0(q)$ as
\begin{eqnarray}
 \chi_0(q)&=&-T\sum_kG(k)G(k+q),
\end{eqnarray}
where the frequency summation is analytically performed in the non-interacting system 
without disorder as
\begin{eqnarray}
\chi_{0}(q)=\sum_{\bf k}\frac{f_{\bf k}-f_{{\bf k}+{\bf q}}}{i\omega_m-\epsilon({\bf k}+{\bf q})
+\epsilon({\bf k})},
\end{eqnarray}
where $f_{\bf k}=\left[\exp((\epsilon({\bf k})-\mu)/T)+1\right]^{-1}$ is the Fermi distributi
on function.
%%%%%%%%%%%%%%%%%%%%%%%%%%%%%%%%%%%%%%%%%%%%%%%%%%%%%%%%%%%%%%%%%%%%%%%%%%%%%%%
\begin{figure}
 \begin{center}
  \includegraphics[scale=0.7]{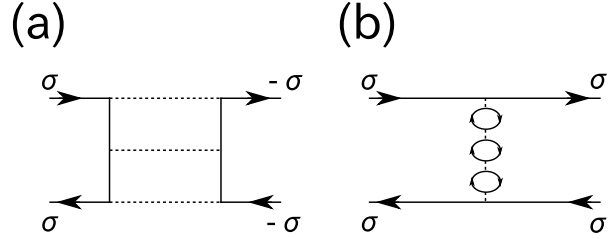}
 \end{center}
 \caption{Maki-Thompson-type vertex correction in RPA.  Broken lines show the on-site interaction 
$U$.  Each graph shows the contribution of (a) the transverse spin fluctuations and 
(b) the longitudinal spin fluctuations and charge fluctuations.}
\label{fig-MT-vertex}
\end{figure}
%%%%%%%%%%%%%%%%%%%%%%%%%%%%%%%%%%%%%%%%%%%%%%%%%%%%%%%%%%%%%%%%%%%%%%%%%%%%%%%

We evaluate the particle-hole irreducible vertex function $\Gamma_1$ representing MT processes and 
AL processes using spin fluctuations within RPA.
The two MT terms in the vertex function are shown in Fig. \ref{fig-MT-vertex}:
\begin{eqnarray}
\Gamma_{1,\rm MT}{ }_{\sigma\sigma'}&\equiv&\sum_{i=a,b}\Gamma^{(i)}_{1,\sigma\sigma'},\\
 \Gamma^{(a)}_{1,\sigma\sigma'}(k,k';q)&=&\frac{U^2\chi_0(k-k')}{1-U\chi_0(k-k')}
\delta_{\sigma,-\sigma'},\\
 \Gamma^{(b)}_{1,\sigma\sigma'}(k,k';q)&=&\frac{U^2\chi_0(k-k')}{1-U^2\chi_0^2(k-k')}
\delta_{\sigma,\sigma'}.
\end{eqnarray}

%%%%%%%%%%%%%%%%%%%%%%%%%%%%%%%%%%%%%%%%%%%%%%%%%%%%%%%%%%%%%%%%%%%%%%%%%%%%%%%
\begin{figure}
 \begin{center}
  \includegraphics[scale=0.7]{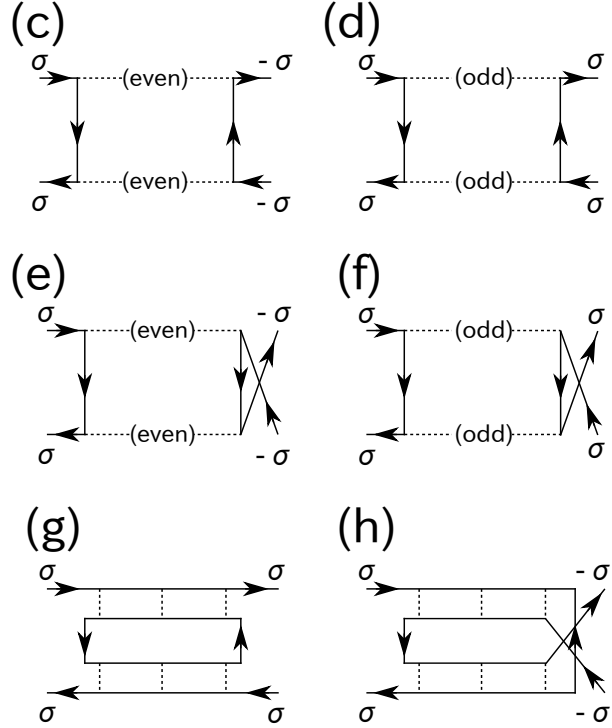}
 \end{center}
 \caption[AL-type vertex correction in random phase approximation.]
{AL-type vertex correction in random phase approximation. Even (odd) denotes the number of 
particle-hole bubbles $\chi_0$.}
\label{fig-AL-vertex}
\end{figure}
%%%%%%%%%%%%%%%%%%%%%%%%%%%%%%%%%%%%%%%%%%%%%%%%%%%%%%%%%%%%%%%%%%%%%%%%%%%%%%%
Six AL terms shown in Fig. \ref{fig-AL-vertex} are given as follows:
\begin{eqnarray}
\Gamma_{1,\rm AL}{ }_{\sigma\sigma'}&\equiv&\sum_{i=c,\cdots h}\Gamma^{(i)}_{1,\sigma\sigma'},\\
 \Gamma^{(c)}_{1,\sigma\sigma'}(k,k';q)&=&T\sum_{q'}\left[
\left(\frac{U}{1-U^2\chi_0^2(q')}\right)
\left(\frac{U}{1-U^2\chi_0^2(q-q')}\right)-U^2\right]G(k+q')G(k'+q')\delta_{\sigma,-\sigma'},\nonumber\\\\
 \Gamma^{(d)}_{1,\sigma\sigma'}(k,k';q)&=&T\sum_{q'}
\left(\frac{U^2\chi_0(q')}{1-U^2\chi_0^2(q')}\right)\left(\frac{U^2\chi_0(q-q')}
{1-U^2\chi_0^2(q-q')}\right)G(k+q')G(k'+q')\delta_{\sigma,\sigma'},\\
 \Gamma^{(e)}_{1,\sigma\sigma'}(k,k';q)&=&T\sum_{q'}\left[
\left(\frac{U}{1-U^2\chi_0^2(q')}\right)\left(\frac{U}
{1-U^2\chi_0^2(q-q')}\right)-U^2\right]G(k+q')G(k'+q-q')\delta_{\sigma,-\sigma'},\nonumber\\\\
 \Gamma^{(f)}_{1,\sigma\sigma'}(k,k';q)&=&T\sum_{q'}
\left(\frac{U^2\chi_0(q')}{1-U^2\chi_0^2(q')}\right)\left(\frac{U^2\chi_0(q-q')}
{1-U^2\chi_0^2(q-q')}\right)G(k+q')G(k'+q-q')\delta_{\sigma,-\sigma'},\\
 \Gamma^{(g)}_{1,\sigma\sigma'}(k,k';q)&=&T\sum_{q'}\left[
\left(\frac{U}{1-U\chi_0(q')}\right)\left(\frac{U}
{1-U\chi_0(q-q')}\right)-U^2\right]G(k+q')G(k'+q')\delta_{\sigma,\sigma'},\\
 \Gamma^{(h)}_{1,\sigma\sigma'}(k,k';q)&=&T\sum_{q'}\left[
\left(\frac{U}{1-U\chi_0(q')}\right)\left(\frac{U}
{1-U\chi_0(q-q')}\right)-U^2\right]G(k+q')G(k'+q-q')\delta_{\sigma,\sigma'}.\nonumber\\
\end{eqnarray}
The second-order diagram in $U$ has been subtracted from $\Gamma^{(i=c, \cdots, h)}_1$ for 
the purpose of avoiding double counting of the MT term $\Gamma_{1,\rm MT}$.  The AL terms consist 
of the particle-hole bubbles for $\Gamma_1^{(i=c, \cdots, f)}$, which is the contribution of the 
longitudinal spin fluctuations and appears when two propagating fluctuations have the same parity 
with respect to the number of particle-hole bubbles, and the particle-hole ladder for 
$\Gamma_{1}^{(i=g,h)}$, which represents the contributions from the transverse spin fluctuations.   
Their total irreducible vertex function is expressed as 
\begin{eqnarray}
 \Gamma_{1,\sigma\sigma'}(k,k';q)=U\delta_{\sigma,-\sigma'}+\Gamma_{1,\rm MT}{ }_{\sigma\sigma'}
(k,k';q)+\Gamma_{1,\rm AL}{ }_{\sigma\sigma'}(k,k';q).\label{eq:total-gamma}
\end{eqnarray}

%\section{Vertex correction to charge susceptibility}
We numerically investigate the role of the MT and AL terms for the charge susceptibility.  
To this end, we calculate the total charge susceptibility using the relation 
\begin{eqnarray}
 \chi_{\rm c}(q)=2\chi_{\rm 0}(q)-2U\chi_{\rm 0}^2(q)+\Delta\chi_{\rm MT}(q)+\Delta\chi_{\rm AL}(q),
\end{eqnarray}
where the third and fourth terms are respectively defined as 
\begin{eqnarray}
\Delta\chi_{\rm MT}(q)&\equiv &T^{2}\sum_{k,k',\sigma,\sigma'}G_{\sigma}(k)G_{\sigma}(k+q)
\Gamma_{1,\rm MT}{ }_{\sigma\sigma'}(k,k';q)G_{\sigma'}(k')G_{\sigma'}(k'+q),\\
\Delta\chi_{\rm AL}(q)&\equiv &T^{2}\sum_{k,k',\sigma,\sigma'}G_{\sigma}(k)G_{\sigma}(k+q)
\Gamma_{1,\rm AL}{ }_{\sigma\sigma'}(k,k';q)G_{\sigma'}(k')G_{\sigma'}(k'+q).
\end{eqnarray}
For the computation of the MT- and AL-type contributions to the susceptibility, the fast Fourier 
transformation (FFT) algorithm is very useful from the viewpoint of numerical efficiency.  
The application method of 
FFT to the AL term is summarized in Appendix B.  We perform the calculations under the condition 
of $32\times32$ lattice sites with 512 Matsubara frequencies.

%%%%%%%%%%%%%%%%%%%%%%%%%%%%%%%%%%%%%%%%%%%%%%%%%%%%%%%%%%%%%%%%%%%%%%%%%%%%%%%
\begin{figure}
 \begin{center}
  \includegraphics[scale=0.5]{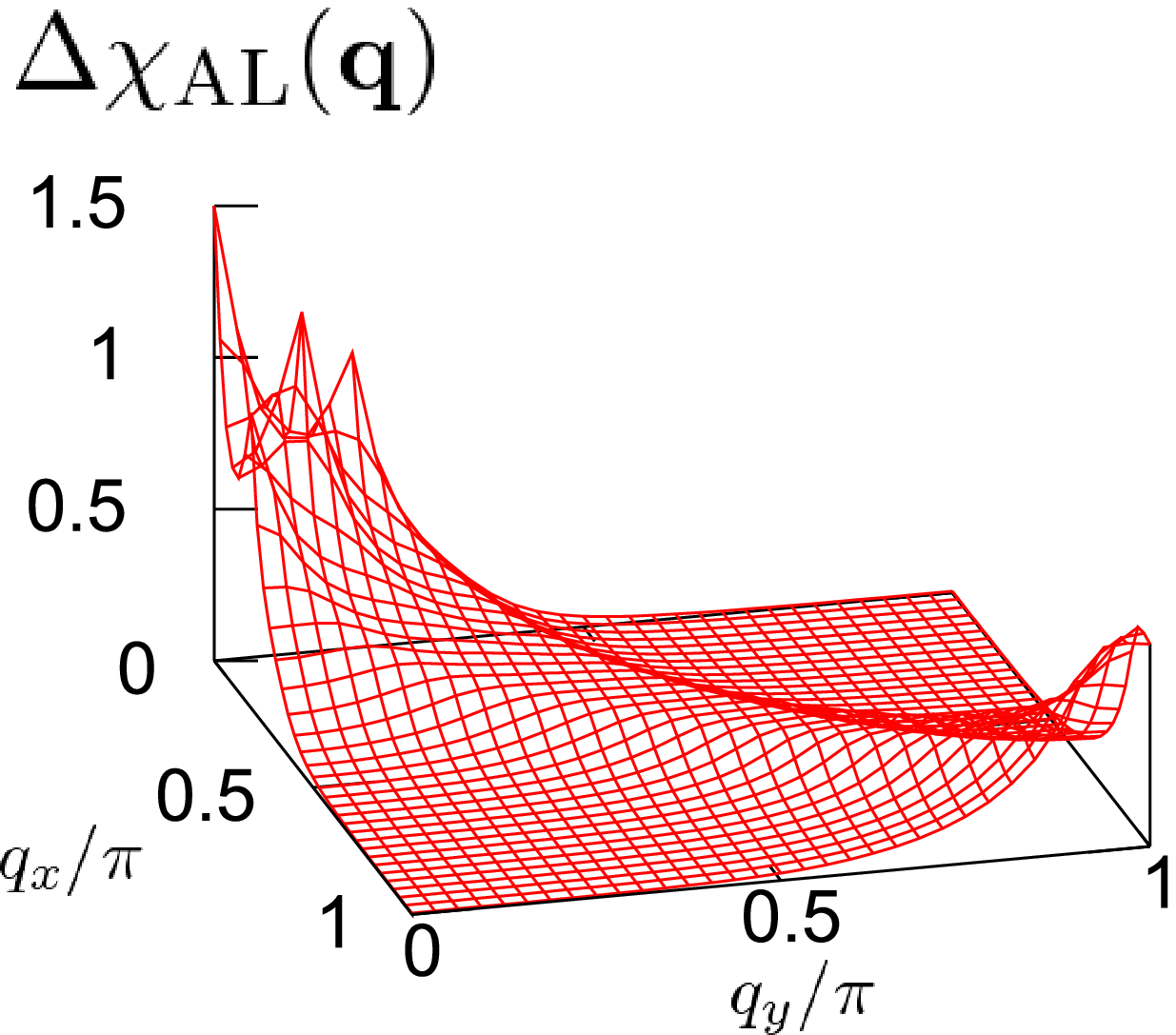}
  \includegraphics[scale=0.5]{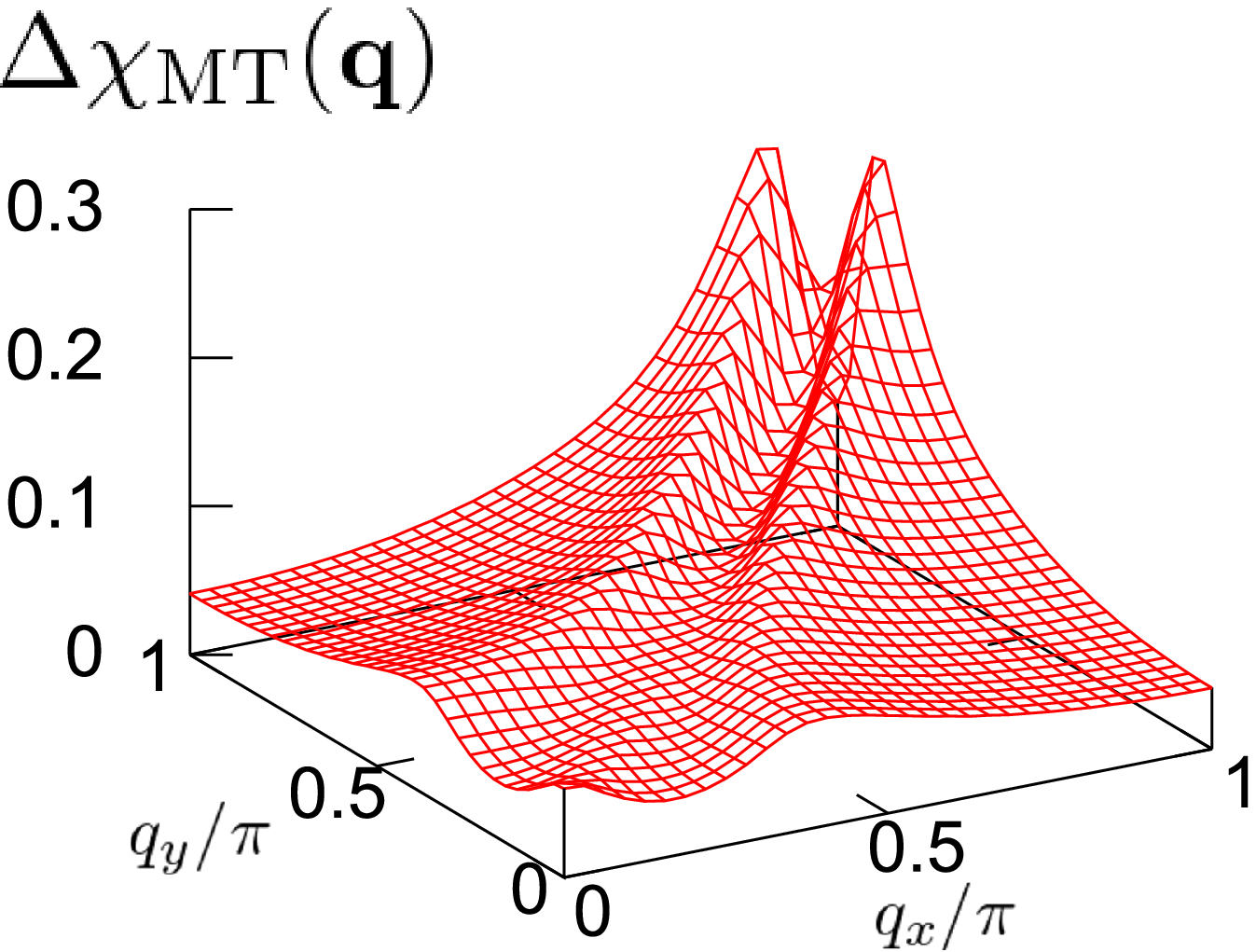}
  \includegraphics[scale=0.5]{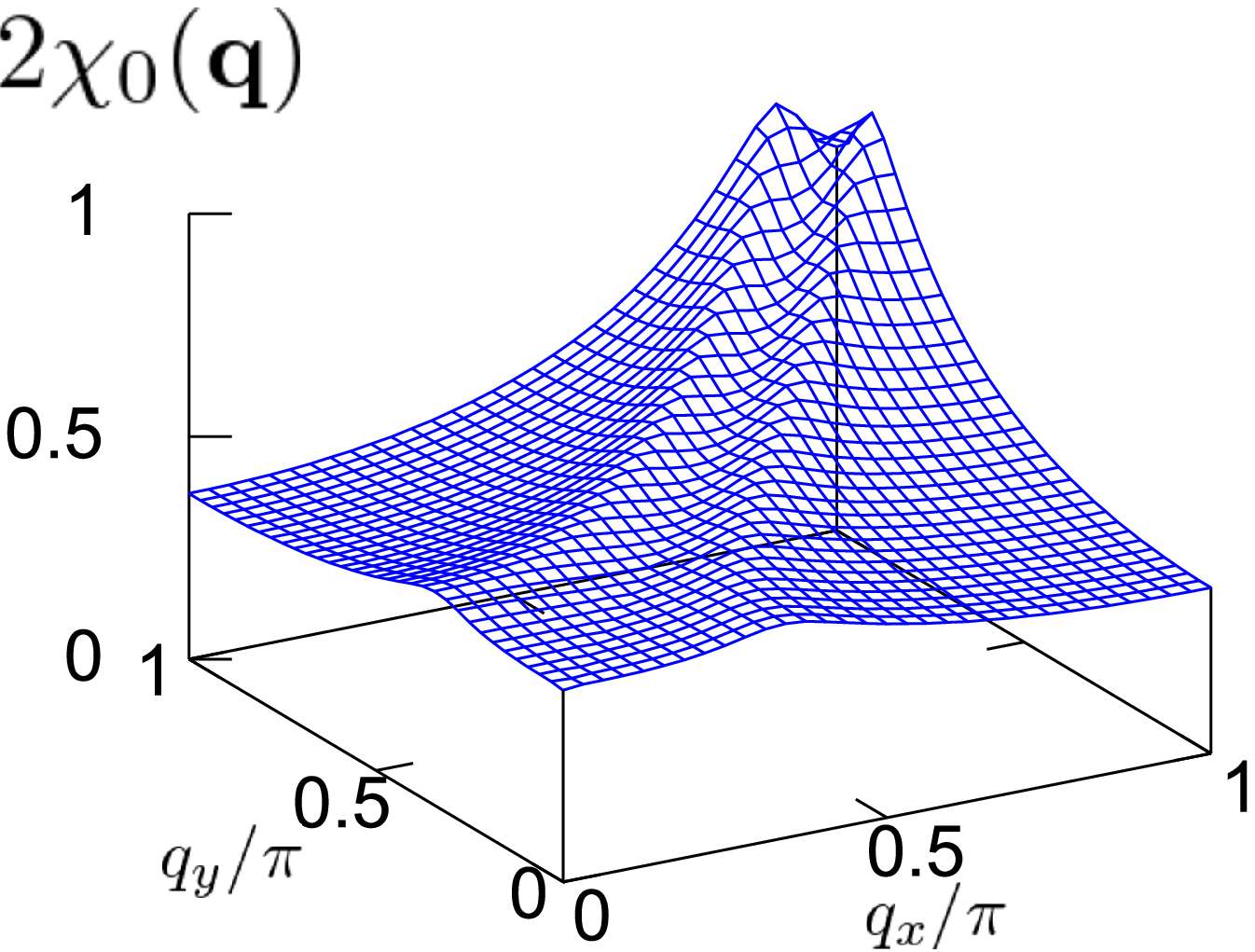}
 \end{center}
 \caption{${\bf q}$ dependence of the Aslamazov Larkin term $\Delta\chi_{\rm AL}({\bf q})$, 
the MT term $\Delta\chi_{\rm MT}({\bf q})$, and the noninteracting part $2\chi_0({\bf q})$ in the static 
charge susceptibility for a parameter set $U=2t$, $n=0.875$, $T=0.05t$, and 
$1-U\chi_0({\bf Q}^*,0)=0.015$.}
\label{fig-chiam0-rpa}
\end{figure}
%%%%%%%%%%%%%%%%%%%%%%%%%%%%%%%%%%%%%%%%%%%%%%%%%%%%%%%%%%%%%%%%%%%%%%%%%%%%%%%
Figure \ref{fig-chiam0-rpa} shows the ${\bf q}$ dependence of each term in the static limit 
($\omega_m$=0). There are several peaks in the AL term $\Delta\chi_{\rm AL}({\bf q})$,  as discussed 
in \S\ref{sec:3}. The maximum is attained at ${\bf q}=0$. The positions of the peaks in the MT 
term $\Delta\chi_{\rm MT}$ almost coincide with the incommensurate ordering vector ${\bf Q}^*_{i}$, 
peak positions of $\chi_0(q)$.
%%%%%%%%%%%%%%%%%%%%%%%%%%%%%%%%%%%%%%%%%%%%%%%%%%%%%%%%%%%%%%%%%%%%%%%%%%%%%%%
\begin{figure}
 \begin{center}
  \includegraphics[scale=0.5]{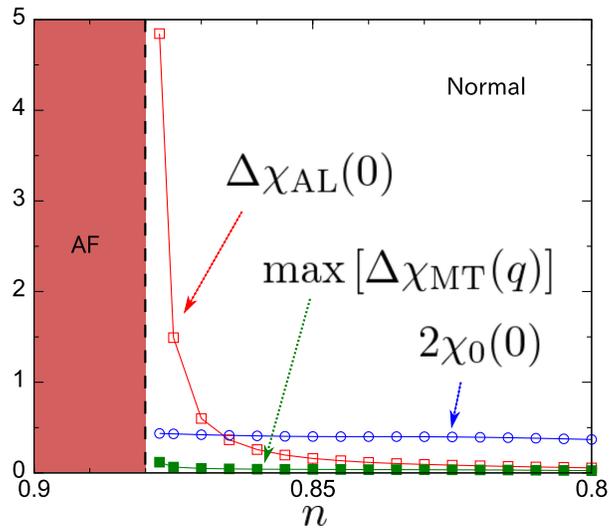}
 \end{center}
 \caption{(Color on line) Doping dependence of the AL term (red), maximum of MT term (green), and 
non-interacting part (blue) in the charge susceptibility when $U=2t$ and $T=0.05t$.  
AF denotes the AF-ordered region.}
\label{fig-ndep-rpa}
\end{figure}
%%%%%%%%%%%%%%%%%%%%%%%%%%%%%%%%%%%%%%%%%%%%%%%%%%%%%%%%%%%%%%%%%%%%%%%%%%%%%%%
Figure \ref{fig-ndep-rpa} shows the doping dependence of each term.
$\Delta\chi_{\rm AL}$ rapidly develops toward the AF transition, as predicted in \S3. In contrast, 
$\Delta\chi_{\rm MT}$ gives only a small contribution to $\chi_c$ even at ${\bf q}\neq 0$.

The singularity in the AL term $\Delta\chi_{\rm AL}$, discussed so far,
arises through the divergence of $m^*$ in eq. (\ref{eq:kappa-landau}).
Hereafter, we show that the repetition of the AL-type vertex correction can give the enhancement 
of the factor $(1+F_0^s)^{-1}$.  The actual calculation for solving the Bethe-Salpeter 
equation based on the irreducible vertex is generally difficult for arbitrary diagrams, 
but we can evaluate the summation of several terms 
in the lowest-order terms.
To check the higher order effects in the MT and AL terms, we consider the following correction in 
the charge susceptibility summed up to the $n$-th order of the irreducible vertex:
\begin{eqnarray}
 \Delta\chi_{{\rm MT},n}&\equiv&GG\Gamma_{1,\rm MT}GG+\cdots +GG\left(\Gamma_{1,\rm MT}GG\right)^n,\\
 \Delta\chi_{{\rm AL},n}&\equiv&GG\Gamma_{1,\rm AL}GG+\cdots +GG\left(\Gamma_{1,\rm AL}GG\right)^n.
\end{eqnarray}
We do not show the higher-order terms including both $\Gamma_{1,\rm MT}$ and $\Gamma_{1,\rm AL}$ 
for simplicity.

%%%%%%%%%%%%%%%%%%%%%%%%%%%%%%%%%%%%%%%%%%%%%%%%%%%%%%%%%%%%%%%%%%%%%%%%%%%%%%%
\begin{figure}
 \begin{center}
  \includegraphics[scale=0.4]{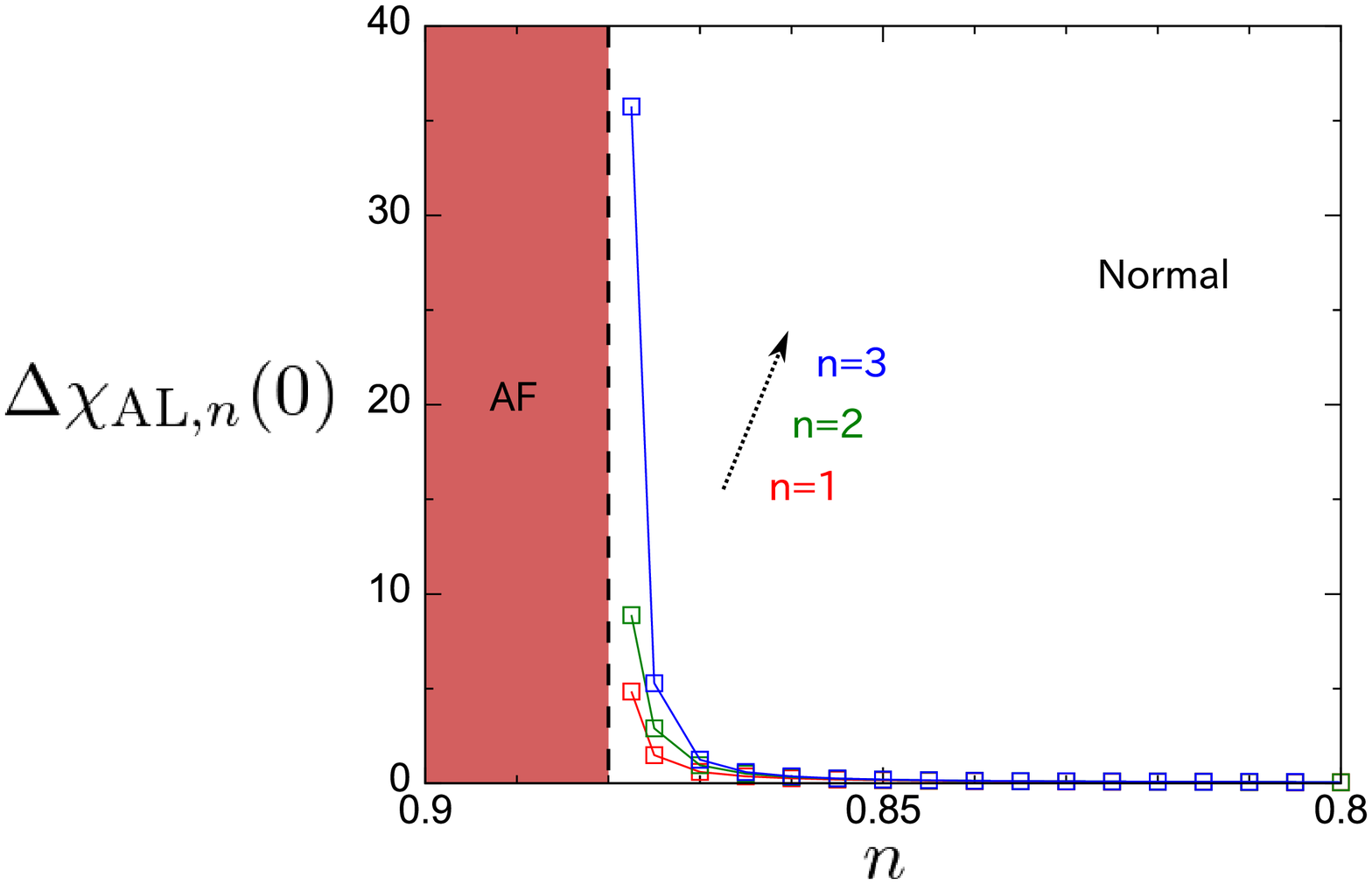}
  \includegraphics[scale=0.4]{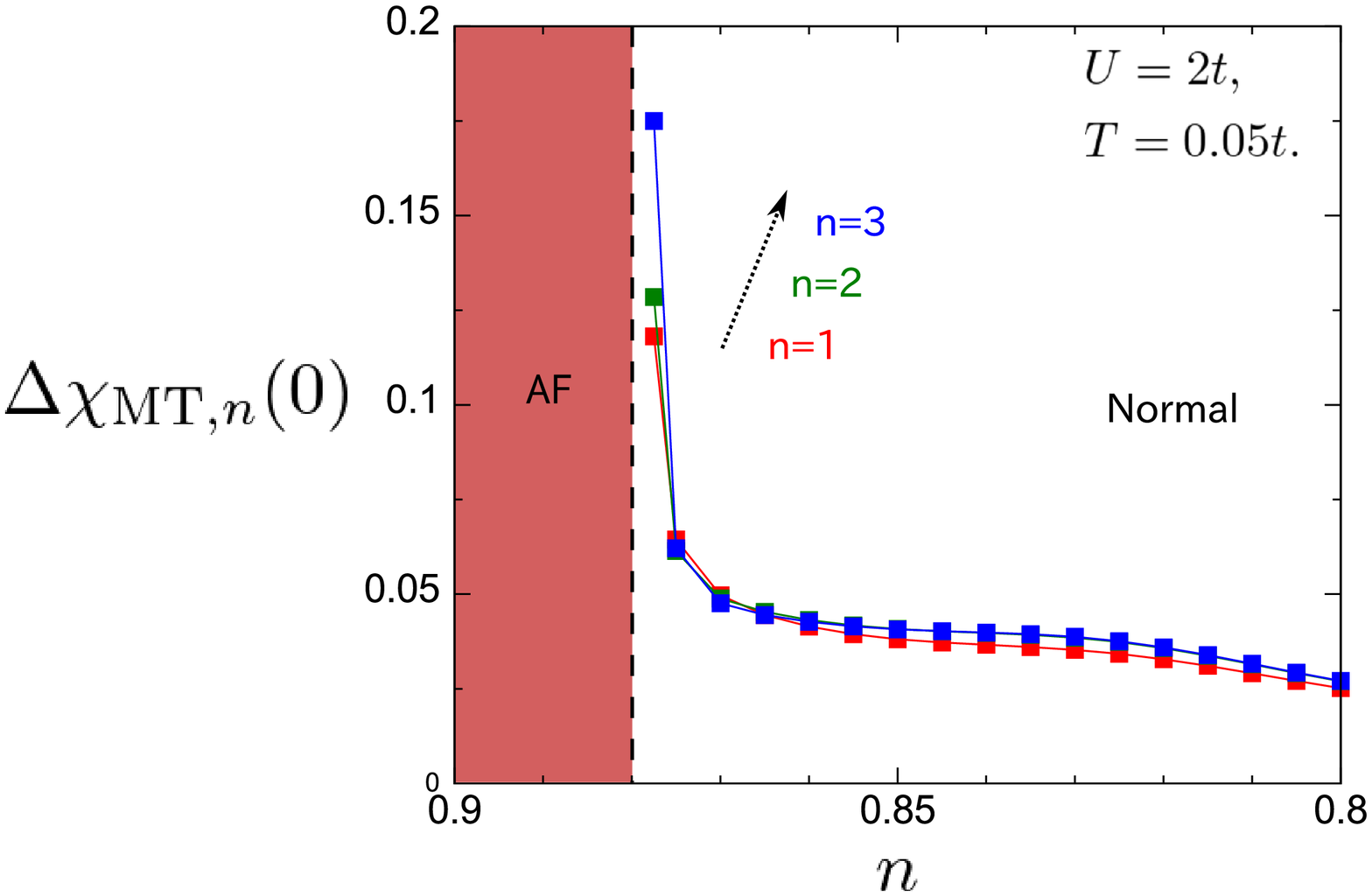}
 \end{center}
 \caption{(Color on line) Higher-order effects on AL term (upper panel) and MT term (lower panel) 
for the charge susceptibility at ${\bf q}=0$. Both terms are evaluated for the parameter set 
$U=0.05t$ and $T=0.05t$.}
\label{fig-high-rpa}
\end{figure}
%%%%%%%%%%%%%%%%%%%%%%%%%%%%%%%%%%%%%%%%%%%%%%%%%%%%%%%%%%%%%%%%%%%%%%%%%%%%%%%
Let us see the tendency of the behavior in the $n\rightarrow\infty$ limit of 
$\Delta\chi_{{\rm AL}, n}$.  Then, $\Delta\kappa_{\rm AL}$ in eq. (\ref{eq:AL-finiteT}) 
and eq. (\ref{eq:AL-zeroT}) in 
two dimensions are replaced with a geometric series\cite{Miyake1994}:
\begin{eqnarray}
 \Delta\kappa_{\rm AL}\rightarrow\Delta\chi_{\rm AL,\infty}(0)=
\begin{cases}
\Delta\kappa_{\rm AL}/\left[1+\frac{\gamma_4}{\pi A}\cdot\frac{N_{\rm F}^2T}{\eta_{\rm s}}\right],
&(T\gg\eta_{s}/2C\pi)\\
\Delta\kappa_{\rm AL}/\left[1+\frac{\gamma_4 N_{\rm F}^2}{\pi AC}\cdot 
\log\frac{Aq_{\rm c}}{\eta_{\rm s}}\right],
&(T=0)
\end{cases}
\end{eqnarray}
where $\gamma_4$ is the quartic mode-mode coupling constant of spin fluctuations. 
$\gamma_4$ for the quartic coupling among four fluctuation modes near the 
${\bf q}\sim {\bf Q}^*$ is expressed as 
\begin{eqnarray}
 \gamma_4=\lambda^4T\sum_{{\bf p},n}G^2({\bf p},i\epsilon_n)G^2({\bf p}+{\bf Q}^*,i\epsilon_n).
\end{eqnarray}
$\gamma_4$ takes a negative value at $T<0.524|\mu|$ in the case of a two-dimensional square 
lattice (see Appendix C).  At that temperature, $\delta\kappa_{\rm AL}$ is enhanced in a 
region far from the AF critical point. 

Figure \ref{fig-high-rpa} shows the behaviors of $\Delta\chi_{{\rm MT},n}$ and 
$\Delta\chi_{{\rm AL},n}$ for $n=1$, $2$, and $3$.  
The singularity in $\Delta\chi_{\rm AL}(0)$ increases as the order $n$ of the diagrams 
increases. These results show that AL-type vertex corrections give a negative value to the 
Landau parameter $F_0^{\rm s}$ as well as $\gamma_4<0$ around the AF criticality in the 
low-temperature region. Concerning the MT term, 
 $\Delta\chi_{\rm MT}$ does not show a remarkable change with higher-order terms.  
The higher-order terms in the MT terms are negligible, but those in the AL terms 
considerably increase the charge susceptibility around the AF criticality.

Note that, throughout this section, we have evaluated the anomaly in the charge compressibility 
with the use of the Green function in a noninteracting system.
 Namely, the self-energy correction arising from electron-electron
 interaction in the Green function is not considered. 
To improve this point, it is desirable to take self-energy correction consistent with 
vertex correction into account so as to satisfy the conservation law.\cite{Baym1961}
However, conserving approximation such as fluctuation exchange
approximation\cite{Bickers1989} do not correctly produce a coherent part in the
single particle Green function and is also not capable of describing the Mott transition.  
Since the singular charge compressibility originates from the formation of the gap in single 
particle excitation at half-filling, self-energy correction has to reproduce the appearance 
of the charge gap.
However, it is still challenging to construct such a conserving theory.
For this difficulty, we ignore self-energy correction for simplicity and consider only 
vertex correction to the impurity potential in the next section.  

\section{Disappearance of coexistence of antiferromagnetism and superconductivity due to large 
impurity potential enhanced by AF fluctuations}
%Discussion: What type of theories are desirable? 
In the remaining sections, it is shown that an inhomogeneity leads to the suppression of both 
the N\'{e}el temperature $T_{\rm N}$ and the SC transition temperature $T_{\rm c}$.  
To this end, we extend the theory of the charge susceptibility taking the effective 
mass enhancement due to the AL term to the impurity scattering problem in cuprates. 
As discussed in previous sections, we focus on the problem of how the impurity scattering 
potential enhanced by the singular charge fluctuation affects $T_{\rm N}$ and $T_{\rm c}$.
Here, we introduce the two-dimensional Hubbard model with the $d_{x^2-y^2}$-wave pairing 
interaction $U_d$ and the impurity potential $v_i$:
\begin{eqnarray}
H&=&\sum_{{\bf k}\sigma}\epsilon({\bf k})c^{\dagger}_{{\bf k}\sigma}c_{{\bf k}\sigma}
+U\sum_{i}n_{i\uparrow}n_{i\downarrow}\nonumber\\
& &-U_d\sum_{k}\phi({\bf k})\phi({\bf k}')
c^{\dagger}_{{\bf k}\uparrow}c^{\dagger}_{-{\bf k}\downarrow}
c_{-{\bf k}'\downarrow}c_{{\bf k}'\uparrow}
+\sum_{i\sigma}v_in_{i\sigma},\label{eq:exhubbard}
\end{eqnarray}
where the form of the gap function $\phi({\bf k})=\cos k_x-\cos k_y$ is chosen to be 
consistent with observations in cuprates. 
The bare impurity potential is given by $v_i=v_0$ for the impurity sites, and $v_i=0$ 
for the other sites.  For La$_{2-\delta}$Sr$_{\delta}$CuO$_4$, it corresponds to 
\begin{eqnarray}
 v_i=
\begin{cases}
 v_0& \text{at CuO$_2$ sites adjacent to Sr},\\
 0& \text{at CuO$_2$ sites adjacent to La}.
\end{cases}
\end{eqnarray}
Namely, we consider a situation where the impurity concentration and the doping rate $\delta$ take 
the same value.  Of course, there exist other types of inhomogeneities, such as the inhomogeneity 
of hopping integrals and the inhomogeneity of interactions on the CuO$_2$ plane, even if they are 
caused by the out-of-plane disorder. However, we assume here the local impurity potential 
in order to study the qualitative aspect of impurity potential renormalization.

%%%%%%%%%%%%%%%%%%%%%%%%%%%%%%%%%%%%%%%%%%%%%%%%%%%%%%%%%%%%%%%%%%%%%%%%%%%%%%%
\begin{figure}
 \begin{center}
  \includegraphics[scale=1]{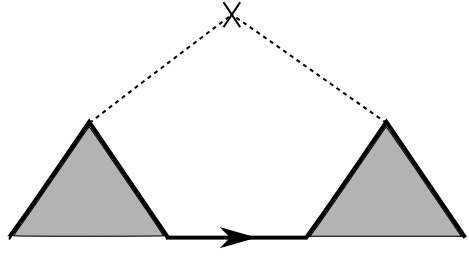}
 \end{center}
 \caption{Self-energy diagram within Born approximation with respect to 
renormalized impurity potential. The shaded triangle represents the renormalized impurity potential.}
\label{fig-se-born}
\end{figure}
%%%%%%%%%%%%%%%%%%%%%%%%%%%%%%%%%%%%%%%%%%%%%%%%%%%%%%%%%%%%%%%%%%%%%%%%%%%%%%%

%impurity potential renormalization
Hereafter, we discuss the effects of vertex correction due to AF critical fluctuations 
neglecting vertex corrections from the $d$-wave attraction $U_d$.
We define the renormalized impurity potential $\tilde{v}$ within the first order approximation 
in the irreducible vertex $\Gamma_{1}$ as
\begin{eqnarray}
 \tilde{v}({\bf k},{\bf k}+{\bf q},i\epsilon_n)=
v_0\Lambda({\bf k},{\bf q},i\epsilon_n),\label{eq:v-renorm}
\end{eqnarray} 
with
\begin{eqnarray}
 \Lambda({\bf k},{\bf q},i\epsilon_n)=1+T\sum_{{\bf k}',n'}G({\bf k}',i\epsilon_n')
G({\bf k}'+{\bf q},i\epsilon_n')\Gamma_1({\bf k}',i\epsilon_n',{\bf k},i\epsilon_n,;{\bf q},0).
\end{eqnarray}
Hereafter, we use the irreducible vertex function $\Gamma_{1}$ given by eq. (\ref{eq:total-gamma}), 
which includes magnetic critical fluctuations within the RPA.

%Born
We evaluate the self-energy in the single-particle Green function within the self-consistent Born approximation.  
With the use of the renormalized impurity potential, we obtain the self-energy shown in 
Fig.\ \ref{fig-se-born} as  
\begin{eqnarray}
 \Sigma_{\rm imp}({\bf k},i\epsilon_n)=
c_{\rm imp}\sum_{{\bf k}'}\tilde{v}^2({\bf k},{\bf k}',i\epsilon_n)
G({\bf k}',i\epsilon_n),\label{eq:sg-Born}
\end{eqnarray}
where $c_{\rm imp}$ is the impurity concentration $c_{\rm imp}=\left|1-n\right|$.
Here, we have assumed that the bare impurity potential is a real number so that 
the renormalized potential ${\tilde v}$ satisfies $\tilde{v}({\bf k},{\bf k}',i\epsilon_n)=
\tilde{v}({\bf k}',{\bf k},i\epsilon_n)$. 
Even if the bare potential is real, the renormalized impurity potential can be a complex number 
in general.    
Even so, the imaginary part of the analytically continued self-energy still gives the damping 
rate of the quasi-particle  
as $1/2\tau_{\bf k}=-{\rm Im}\Sigma_{\rm imp}({\bf k},\epsilon+i\delta)$. 

We must comment on the first-order term of $\tilde{v}$ in the self-energy, which is given by
\begin{eqnarray}
 \Delta\Sigma^{(1)}({\bf k},i\epsilon_n)=
c_{\rm imp}\tilde{v}({\bf k},{\bf k},i\epsilon_n).\label{eq:1st-sg}
\end{eqnarray}
For the bare potential, since it has the general form of 
$v_0({\bf k},{\bf k}',i\epsilon)=v_0({\bf k}-{\bf k}')$, the first-order term is constant, 
giving only a chemical potential shift.  Unlike the bare potential, eq. (\ref{eq:1st-sg}) gives 
the ${\bf k}$ and $\epsilon_n$ dependences.  Unfortunately, the sign of the renormalized potential 
${\tilde v}$ in Eq. (\ref{eq:1st-sg}) depends on the sign of the bare potential $v_0$.  
Thus, it shows an anomalous  behavior that the damping rate easily becomes negative.  
However, this term must give only the chemical potential shift in the Green function 
$G$ constituting the self-energy like
\begin{eqnarray}
\Sigma({\bf k},i\epsilon_n)&=&T\sum_{{\bf q},m}
V({\bf q},i\omega_m)G({\bf k}+{\bf q},i\epsilon_n+i\omega_m),
\label{eq:sgfluc}\\
V({\bf q},i\omega_m)&=&U+\frac{3}{2}\frac{U^2\chi_0({\bf q},i\omega_m)}{1-U\chi_0({\bf q},i\omega_m)}+
\frac{1}{2}\frac{U^2\chi_0({\bf q},i\omega_m)}{1+U\chi_0({\bf q},i\omega_m)}-U^2\chi_0({\bf q},
i\omega_m).
\label{eq:sigma-cor}
\end{eqnarray}
This self-energy $\Sigma$ is related to the irreducible vertex function by 
$\frac{\delta\Sigma(k)}{\delta G(k')}=\Gamma_1(k,k';q=0)$.
We do not consider this kind of self-energy correction in the present paper.
However, because eq. (\ref{eq:1st-sg}) should originally be absorbed into eq. (\ref{eq:sgfluc}), 
we eliminate this first-order term.

We note that the Born approximation is valid in the case of a dilute concentration of 
impurities with a weak potential. In addition, since it overestimates the damping rate for a strong 
impurity potential, the application of the $t$-matrix theory is desirable in such a case.  
Nevertheless, here we use the Born approximation for simplicity because better theories, such as 
the $t$-matrix approximation, are numerically difficult to adopt for renormalized impurity 
potential including complex ${\bf k}$ and ${\bf k}'$ dependences.

%Dyson
If we have the calculated self-energy $\Sigma_{\rm imp}({\bf k},i\epsilon_n)$ due to impurities 
in eq. (\ref{eq:sg-Born}), 
the single-particle Green function is obtained from the Dyson equation:
\begin{eqnarray}
 G^{-1}({\bf k},i\epsilon_n)=i\epsilon_n-\epsilon({\bf k})+\mu-
\Sigma_{\rm imp}({\bf k},i\epsilon_n).\label{eq:dyson}
\end{eqnarray}
Using this type of Green function, we repeatedly calculate the vertex function in 
eq. (\ref{eq:total-gamma}), the renormalized impurity potential in eq. (\ref{eq:v-renorm}), 
and the self-energy in eq. (\ref{eq:sg-Born}).  
This self-consistent calculation scheme is performed until $G$,
$\Sigma_{\rm imp}$ and $\Gamma$ converge.
If a self-consistent solution is obtained, we can evaluate the effect of
renormalized impurity scattering on the SC 
transition temperature $T_{\rm c}$ and the N\'{e}el temperature $T_{\rm N}$.
Here, the carrier density $n$ is evaluated as 
\begin{eqnarray}
n&=&2T\sum_{{\bf k},n}G({\bf k},i\epsilon_n)e^{i\epsilon_n\delta}\nonumber\\
&=&1+4T\sum_{{\bf k},n\ge 0}{\rm Re}G({\bf k},i\epsilon_n).\label{eq:carrier}
\end{eqnarray}
We note here that ${\partial n}/{\partial \mu}$ calculated from the
carrier density in eq. (\ref{eq:carrier}) and from the Green function in 
eq. (\ref{eq:dyson}) does not give divergence toward AF
criticality because $G({\bf k},i\epsilon_n)$ does not include the 
self-energy correction of eq. (\ref{eq:sgfluc}) with an effect of critical 
AF fluctuations.  In this sense, relation eq. (\ref{eq:kappa-rel}) is broken due to
difference in the level of perturbation between $\kappa$ calculated from the 
$\mu$ dependences of $n$ and $\chi_{\rm c}$ constructed by using an approximate 
vertex function $\Gamma_{1}$, eq. (\ref{eq:total-gamma}). 

%T_N
The AF transition temperature $T_{\rm N}$ is determined with the use of the converged Green 
function by
\begin{eqnarray}
 1-\alpha&=&U\max\left[\chi_0({\bf Q},0;T=T_{\rm N})\right],\\
\chi_0({\bf Q},0;T)&=&-T\sum_{{\bf k},n}G({\bf k},i\epsilon_n)G({\bf k}+{\bf Q},i\epsilon_n).
\label{eq:51}
\end{eqnarray}
The ordering vector ${\bf Q}$, which gives the maximum $\chi_0({\bf Q},0)$, is not restricted 
to the commensurate AF vector ${\bf Q}=(\pi,\pi)$ in general. Thus, hereafter, our criterion 
furthermore gives an incommensurate spin density wave transition.
The true solution should be given at $\alpha=0$. 
However, this criterion never gives the AF and SDW transition because the AL term gives the 
relation $\eta_{\rm c}=\eta_{\rm s}$ in eq. (\ref{eq:tau}) and the damping rate diverges at 
the true AF critical point $\eta_{\rm s}=\alpha=0$.
This serious problem is solved by the extension tothe $t$-matrix theory in which the damping 
rate saturates at a constant value at $\eta_{\rm c}=0$,  as discussed in \S3. 
Because the extension to the $t$-matrix theory of the renormalized impurity
potential is very difficult to compute within reasonable computation time,
we choose the condition of $\alpha=0.005$ to estimate $T_{\rm N}$.
This $\alpha$ corresponds to the situation where the 2D AF coherence length takes 
$\xi_{\rm AF}\sim 13a$, with $a$ being the lattice constant, at $U=2$ and in the half-filling.

%Tc
The SC transition temperature $T_{\rm c}$ is given by the Thouless criterion
\begin{equation}
 1=U_dT\sum_{{\bf k},n}\left|G({\bf k},i\epsilon_n)\right|^2\phi^2({\bf k}).\label{thouless}
 \end{equation}
Since the impurity effect is included in the dressed Green function, a solution of 
eq. (\ref{thouless}) gives a decrease in $T_{\rm c}$ in disordered systems.

%%%%%%%%%%%%%%%%%%%%%%%%%%%%%%%%%%%%%%%%%%%%%%%%%%%%%%%%%%%%%%%%%%%%%%%%%%%%%%%
\begin{figure}
 \begin{center}
  \includegraphics[scale=0.5]{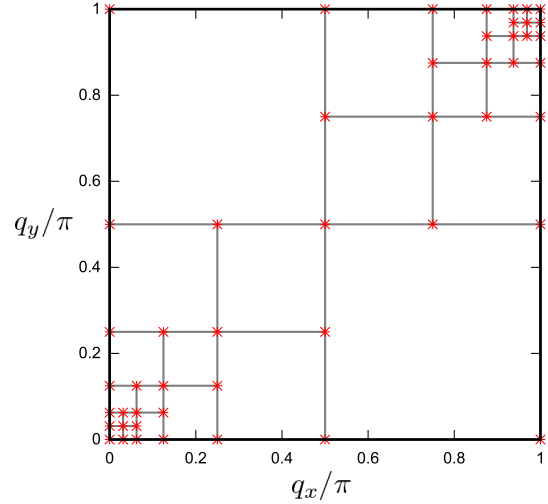}
 \end{center}
 \caption{Mesh in ${\bf q}$-space for 32$\times$32 lattice for numerical calculations. 
The ${\bf q}$-summation in the Born self-energy, eq.\ (\ref{eq:sg-Born2}), is performed over 
the center in the square surrounded by points with X symbols.}
\label{fig-mesh}
\end{figure}
%%%%%%%%%%%%%%%%%%%%%%%%%%%%%%%%%%%%%%%%%%%%%%%%%%%%%%%%%%%%%%%%%%%%%%%%%%%%%%%
%q-sum
The numerical summation in eq. (\ref{eq:sg-Born}) requires a much longer time for calculations 
than that for $\chi_0$, eq. (\ref{eq:51}). Then, we rewrite Eq. (\ref{eq:sg-Born}) as
\begin{eqnarray}
 \Sigma_{\rm imp}({\bf k},i\epsilon_n)=c_{\rm imp}v_0^2\sum_{\bf k}
G({\bf k},i\epsilon_n)+c_{\rm imp}\sum_{{\bf q}}\left(
\tilde{v}^2({\bf k},{\bf k}+{\bf q};i\epsilon_n)-v_0^2
\right)G({\bf k}+{\bf q},i\epsilon_n).\label{eq:sg-Born2}
\end{eqnarray}
The first term corresponds to the Born approximation for the bare impurity potential $v_0$, and its 
summation can be easily performed. In contrast, the second term is numerically more difficult 
to evaluate.  
Here, since the structure of the vertex correction appears at ${\bf q}=0$ and ${\bf q}=(\pi,\pi)$, 
we approximate the summation using ${\bf q}$-points with a high-priority around ${\bf q}=0$ 
and $(\pi,\pi)$.  The resulting mesh points are shown in Fig. \ref{fig-mesh} for the 
$32\times32$ lattice. By discretizing the ${\bf q}$-integral in eq. (\ref{eq:sg-Born2}) with 
the use of the trapezoidal formula in each 
squared region, we can derive the weight of each mesh point in the discretized summation.

Although we take into account only the self-energy owing to impurities in the present paper, 
the above framework enables us to discuss how the mean-field phase diagram is influenced by 
renormalized impurity scattering.

\section{Doping Phase Diagram in High-$T_{\rm c}$ Cuprates with Disorder}
First, let us investigate the behavior of the enhancement factor of the impurity 
potential and the Born self-energy in the limit of $v_0\rightarrow 0$ for a parameter set 
$U=2t$ and $n=0.875$, where $\eta_{\rm s}\equiv 1-U{\rm max}[\chi_0({\bf Q},0)]=0.015$.

%%%%%%%%%%%%%%%%%%%%%%%%%%%%%%%%%%%%%%%%%%%%%%%%%%%%%%%%%%%%%%%%%%%%%%%%%%%%%%%
\begin{figure}
 \begin{center}
  \includegraphics[scale=0.5]{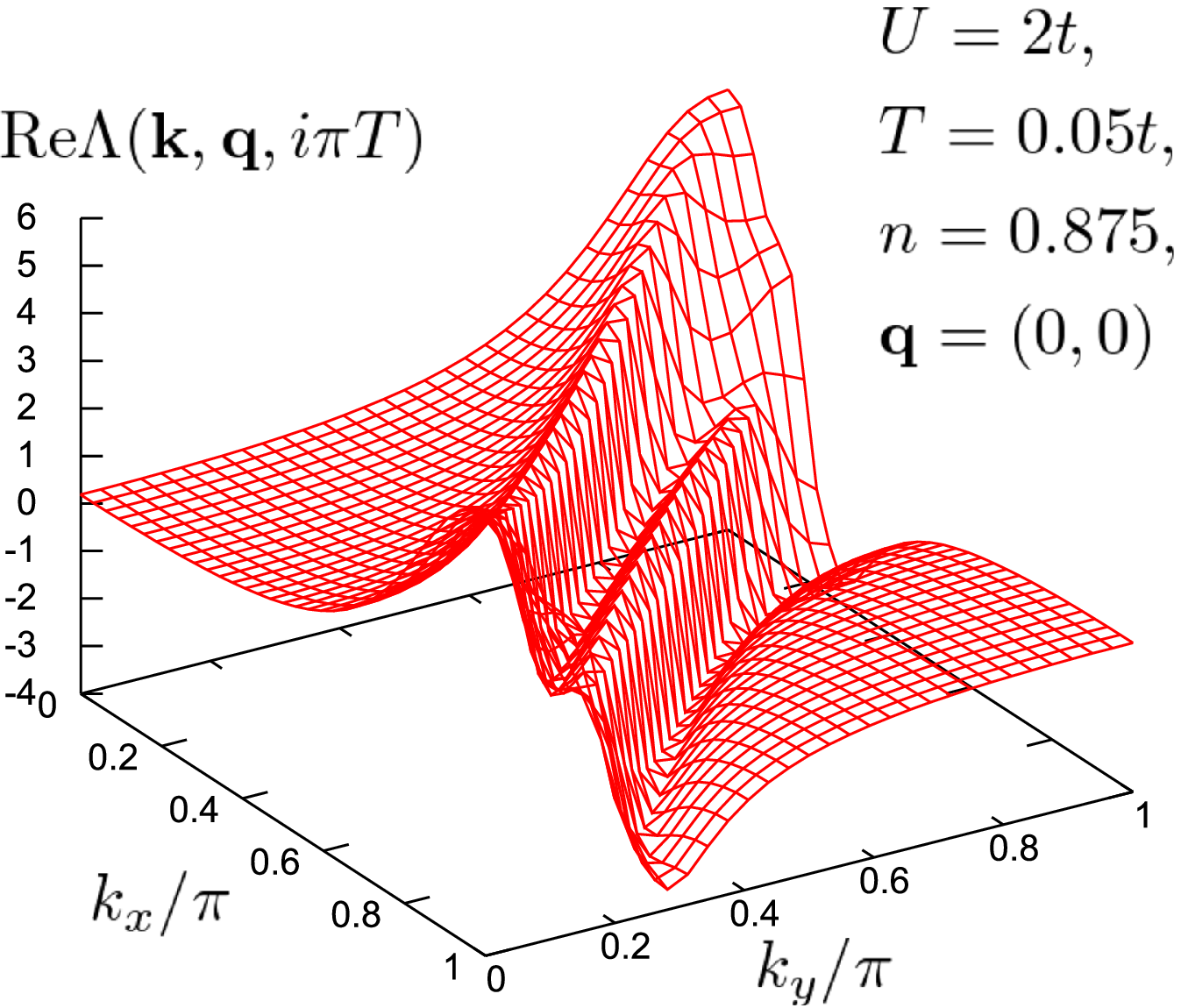}
  \includegraphics[scale=0.5]{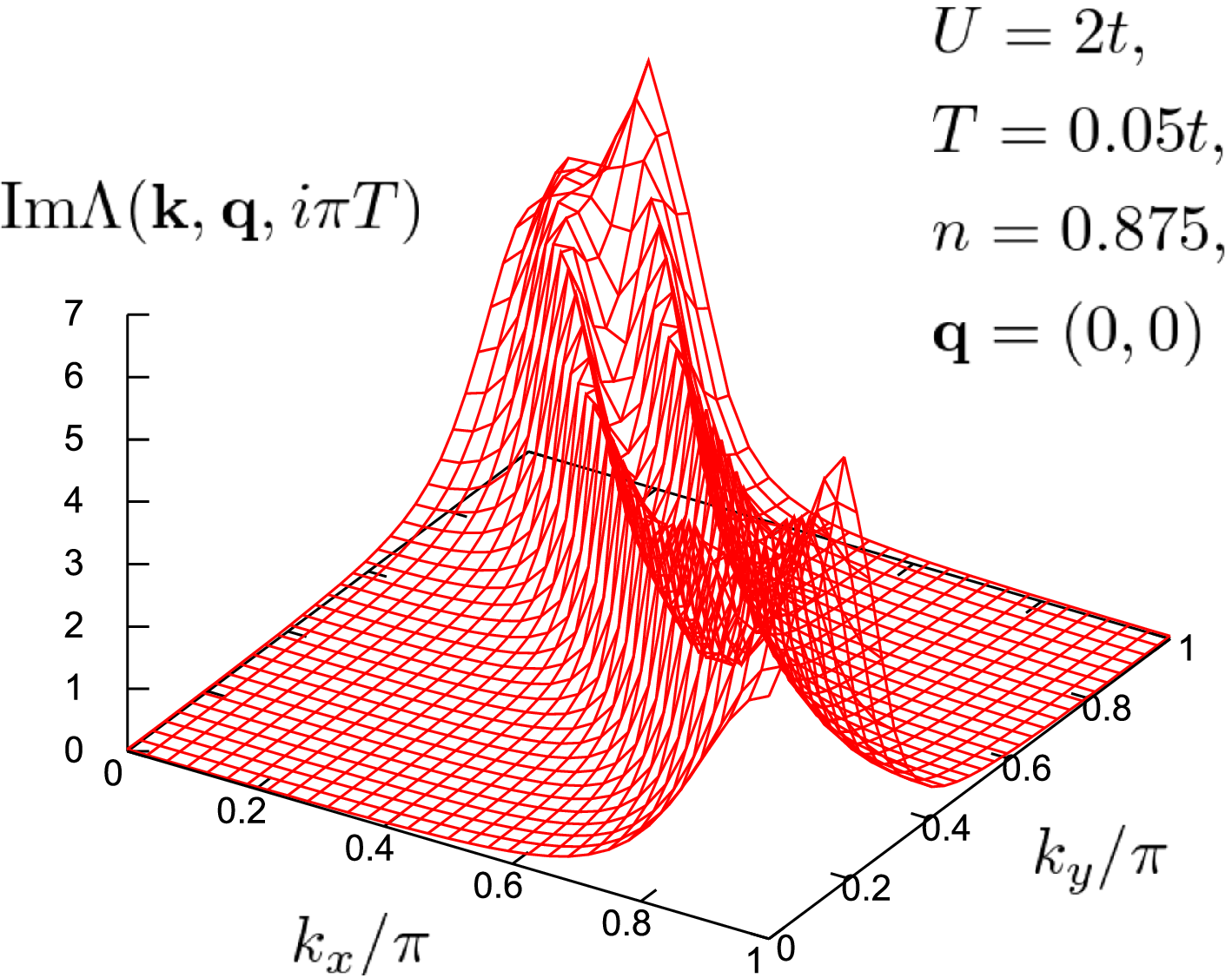}
 \end{center}
 \caption{Real and imaginary parts of the renormalization factor
 $\Lambda({\bf k},{\bf k},i\pi T)$ 
for the impurity potential as a function of the momentum ${\bf k}$ of scattering electron. 
The momentum 
transfer ${\bf q}$ is chosen to be zero.}
\label{fig-lambdak}
\end{figure}
%%%%%%%%%%%%%%%%%%%%%%%%%%%%%%%%%%%%%%%%%%%%%%%%%%%%%%%%%%%%%%%%%%%%%%%%%%%%%%%
Figure \ref{fig-lambdak} shows the incident momentum (${\bf k})$ dependence of the 
renormalization factor of 
the impurity potential $\Lambda({\bf k},{\bf q},i\epsilon_n)$ in the forward scattering 
limit ${\bf q}=0$.  Its real part exhibits a peak at the Fermi surface, but the peak in the 
imaginary part deviates from 
the Fermi surface. Thus, only the enhanced real part contributes to low-energy impurity 
scattering processes.
Note that the renormalization factor in our theory satisfies the relation
$\lim_{{\bf q}\rightarrow 0} \Lambda({\bf k},{\bf q},0+i\delta)=
1+\partial\Sigma({\bf k},0+i\delta)/\partial \mu$, where the self-energy due to the electron
correlation $\Sigma({\bf k},i\epsilon_n)$ is given by
eq. (\ref{eq:sgfluc}). In a normal Fermi liquid, ${\rm Im}\Sigma({\bf
k},\epsilon+i\delta)\propto {\rm max}(\epsilon^2, \pi^2T^2)$
 is satisfied.\cite{AGD1965}
 Thus, $\lim_{{\bf q}\rightarrow 0}{\rm Im}
\Lambda({\bf k},{\bf q},0+i\delta)=\partial {\rm Im}
\Sigma({\bf k},0+i\delta)/{\partial \mu}$
generally takes a nonzero value at finite temperatures.
Therefore, the renormalized impurity potential generally takes a complex value 
even in the static limit ($i\epsilon_n\rightarrow 0+i\delta$), even if the bare potential is real.
For these reasons, the relation 
$\tilde{v}({\bf k},{\bf k}',i\epsilon)=\tilde{v}^*({\bf k}',{\bf k},i\epsilon)$ required in 
the bare potential is not satisfied.
%%%%%%%%%%%%%%%%%%%%%%%%%%%%%%%%%%%%%%%%%%%%%%%%%%%%%%%%%%%%%%%%%%%%%%%%%%%%%%%
\begin{figure}
 \begin{center}
  \includegraphics[scale=0.5]{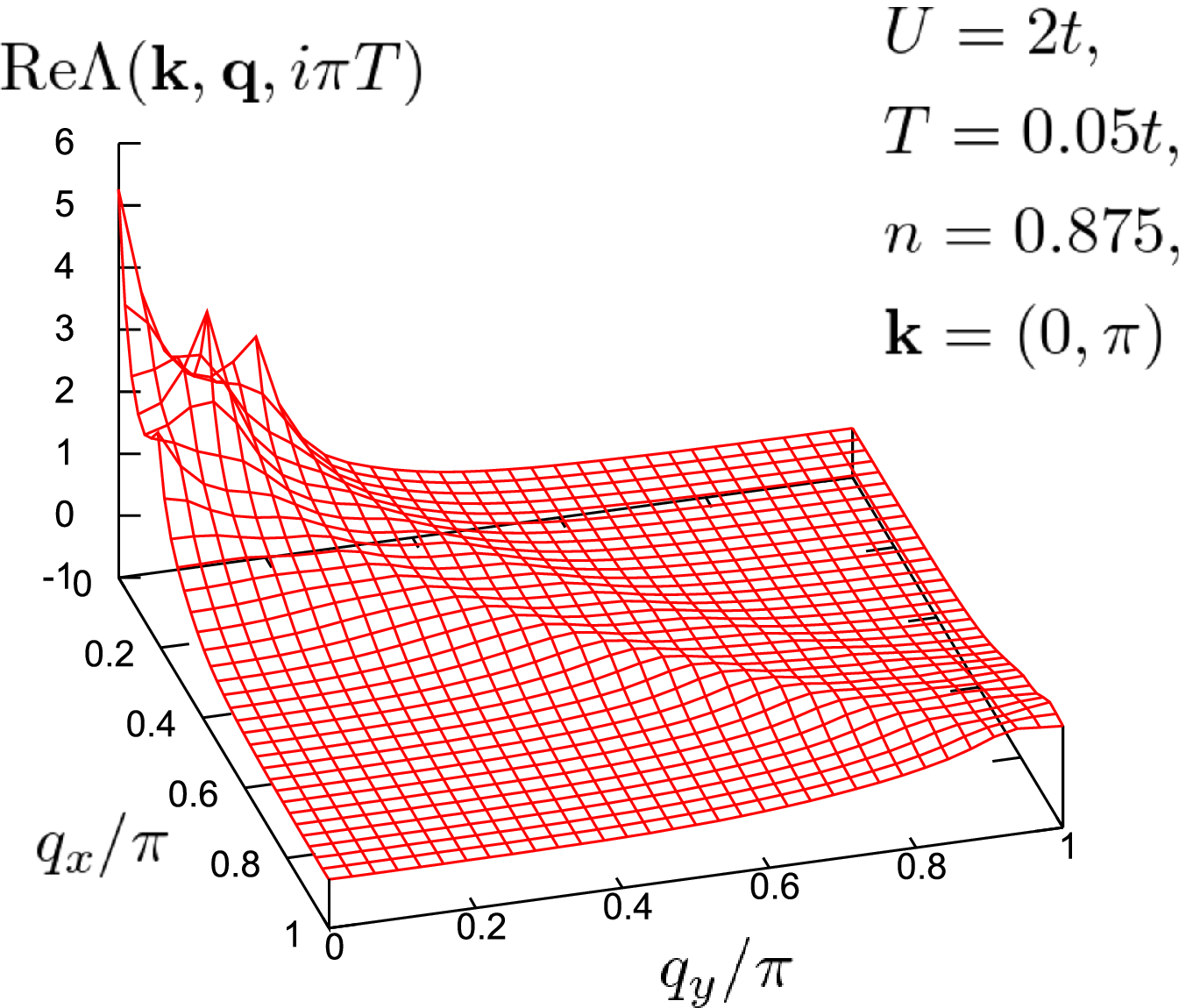}
  \includegraphics[scale=0.5]{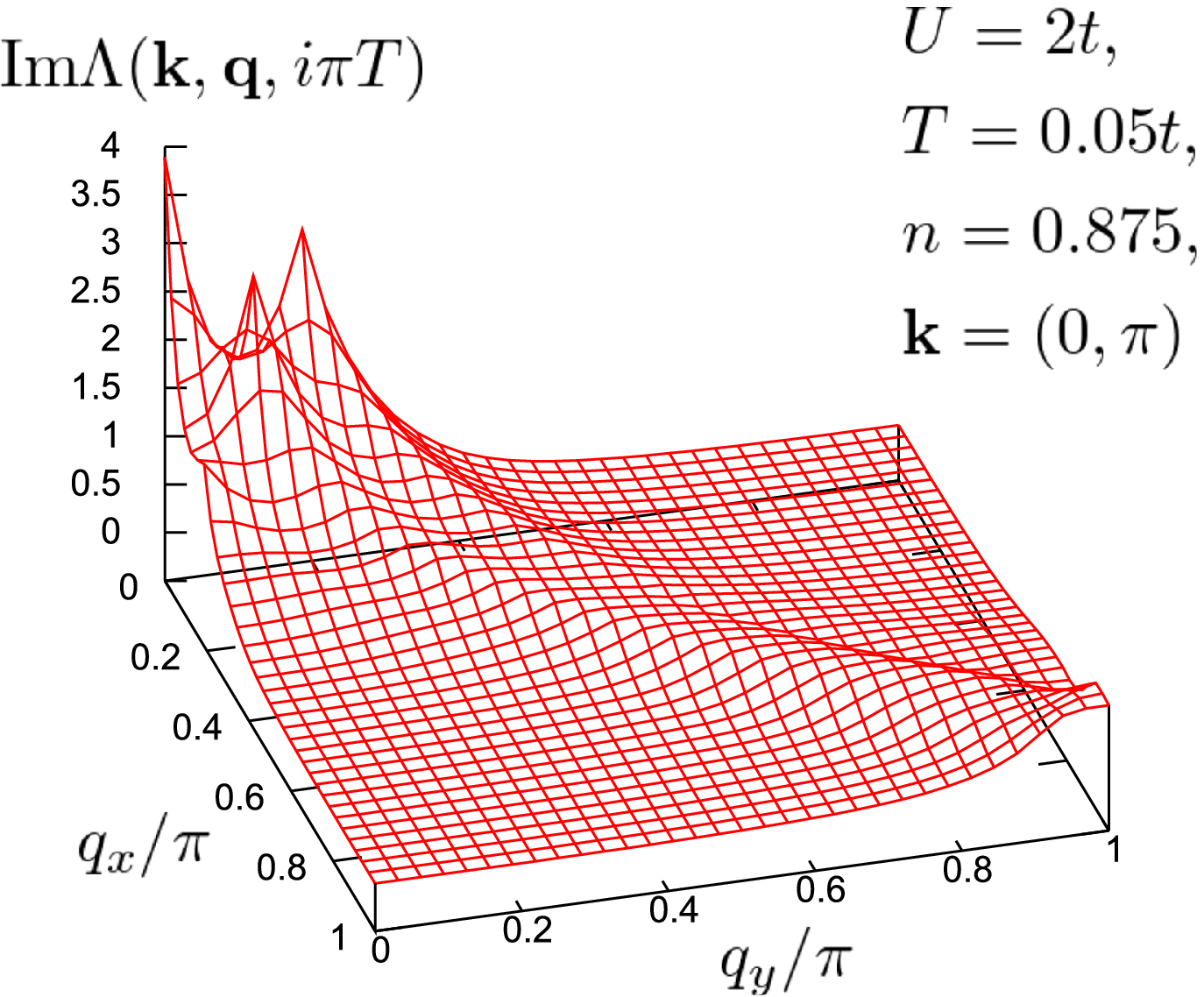}
 \end{center}
 \caption{Real and imaginary parts of impurity potential renormalization factor 
$\Lambda({\bf k},{\bf k}+{\bf q},i\pi T)$ as functions of transfer ${\bf q}$. The incident momentum 
is chosen to be ${\bf k}=(\pi, \pi)$, around which the real part shows its peak.}
\label{fig-lambdaq}
\end{figure}
%%%%%%%%%%%%%%%%%%%%%%%%%%%%%%%%%%%%%%%%%%%%%%%%%%%%%%%%%%%%%%%%%%%%%%%%%%%%%%%
Figure \ref{fig-lambdaq} shows a momentum transfer (${\bf q}$) dependence of the 
renormalization factor.  
Its global maximum appears at ${\bf q}=0$, as in the case of the AL-type correction of the charge 
susceptibility. Secondary peaks are also obtained near ${\bf q}=0$ as the AL term in the charge 
susceptibility.  
%%%%%%%%%%%%%%%%%%%%%%%%%%%%%%%%%%%%%%%%%%%%%%%%%%%%%%%%%%%%%%%%%%%%%%%%%%%%%%%
\begin{figure}
 \begin{center}
  \includegraphics[scale=0.5]{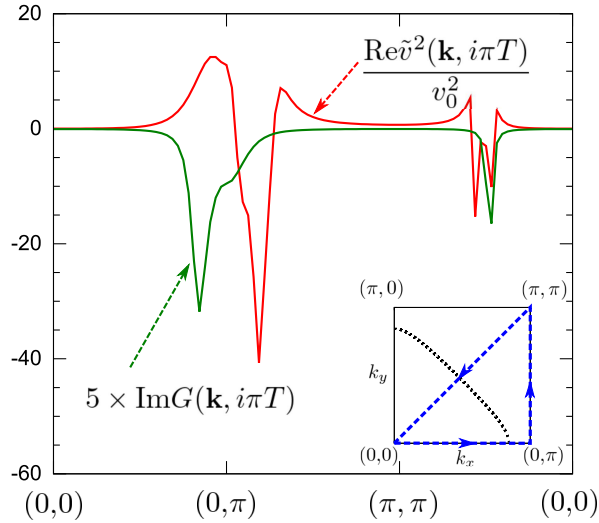}
 \end{center}
 \caption{Comparison of real part of $\tilde{v}^2({\bf k},{\bf k},i\pi T)$ (solid line) and 
Im$G({\bf k},i\pi T)$ (dotted line) in the ${\bf k}$-space at $n=0.875$ and $T=0.05t$. 
The variable {\bf k} sweeps along the broken line with the arrow shown in the inset. 
The Fermi surface for 
this case is drawn with a dotted line in the inset.}
\label{fig-comp-r}
\end{figure}
%%%%%%%%%%%%%%%%%%%%%%%%%%%%%%%%%%%%%%%%%%%%%%%%%%%%%%%%%%%%%%%%%%%%%%%%%%%%%%%
Correspondingly, we show the structure of the integral of the self-energy in eq. (\ref{eq:sg-Born}).
The renormalized impurity potential is mainly enhanced in the forward scattering channel. 
Thus, the Born self-energy $\Sigma_{\rm imp}$ can be approximated near the criticality as
\begin{eqnarray}
 \Sigma_{\rm imp}({\bf k},i\epsilon_n)&\propto& \tilde{v}^2({\bf k},{\bf k},i\epsilon_n)
G({\bf k},i\epsilon_n)\nonumber\\
&\sim&{\rm Re}[\tilde{v}^2({\bf k},{\bf k},i\epsilon_n)]{\rm Im}G({\bf k},i\epsilon_n),
\end{eqnarray}
where we have ignored the contribution of ${\rm Re}G({\bf k},i\epsilon_n)$ in the second 
line, because ${\rm Re}G({\bf k},0+i\delta)$ approaches zero at the Fermi surface 
${\bf k}={\bf k}_{\rm F}$.  
Figure \ref{fig-comp-r} shows the ${\bf k}$ dependence of 
${\rm Re}[\tilde{v}^2({\bf k},{\bf k},i\pi T)]$ and 
${\rm Im}G({\bf k},i\pi T)$. The peak in ${\rm Im}G({\bf k},i\pi T)$ corresponds to the Fermi surface 
momentum.  The square of the renormalized impurity potential $\tilde{v}^2$ takes both positive and 
negative values. However, since it takes a positive value at the Fermi surface in the $(0,\pi)$ 
direction, the scattering rate of quasi-particle 
$1/2\tau_{\bf k} \sim {\rm Im}\Sigma_{\rm imp}({\bf k},i\pi T)$ is largely enhanced at the Fermi surface. 

%%%%%%%%%%%%%%%%%%%%%%%%%%%%%%%%%%%%%%%%%%%%%%%%%%%%%%%%%%%%%%%%%%%%%%%%%%%%%%%
\begin{figure}
 \begin{center}
  \includegraphics[scale=0.5]{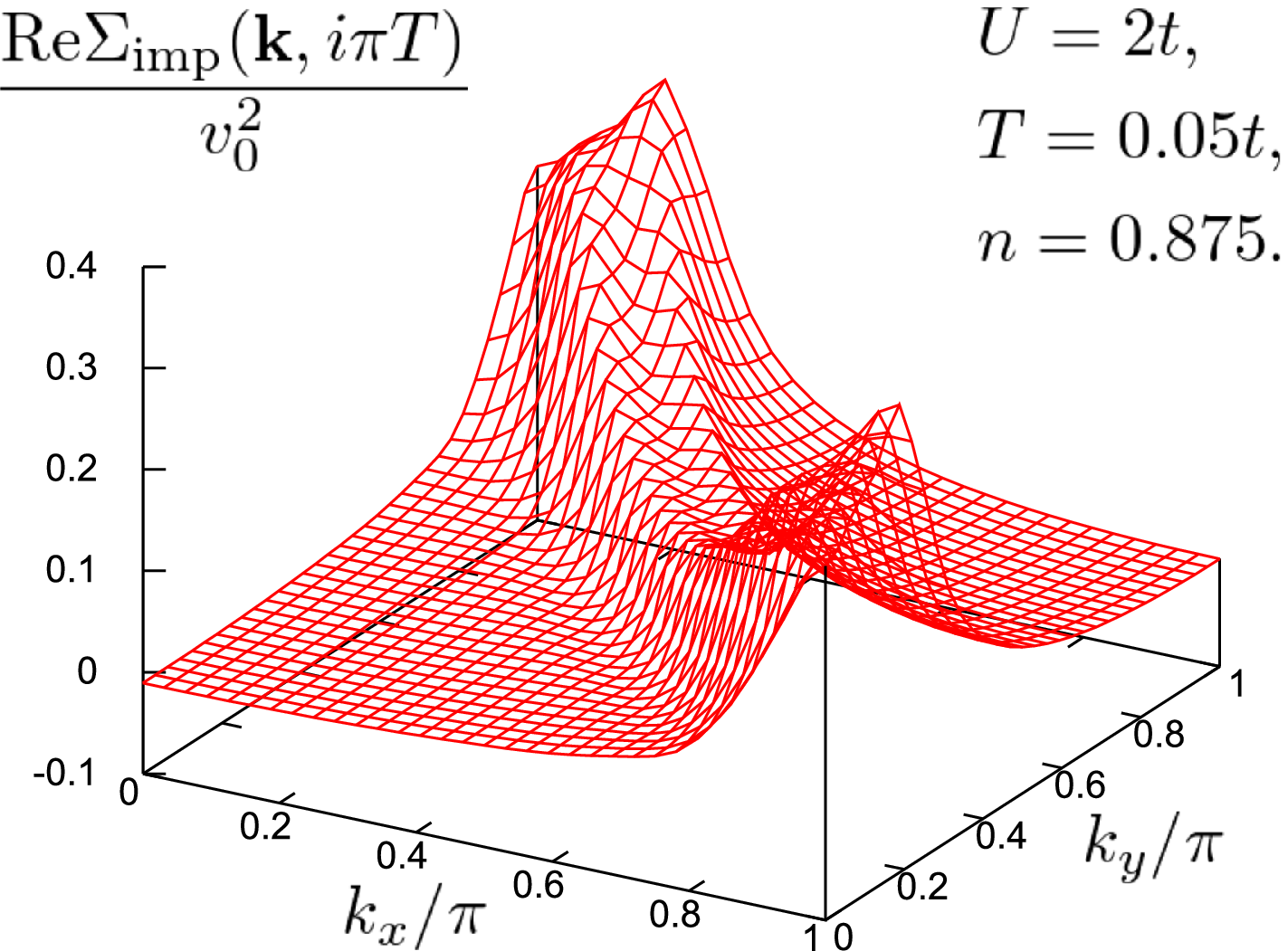}
  \includegraphics[scale=0.5]{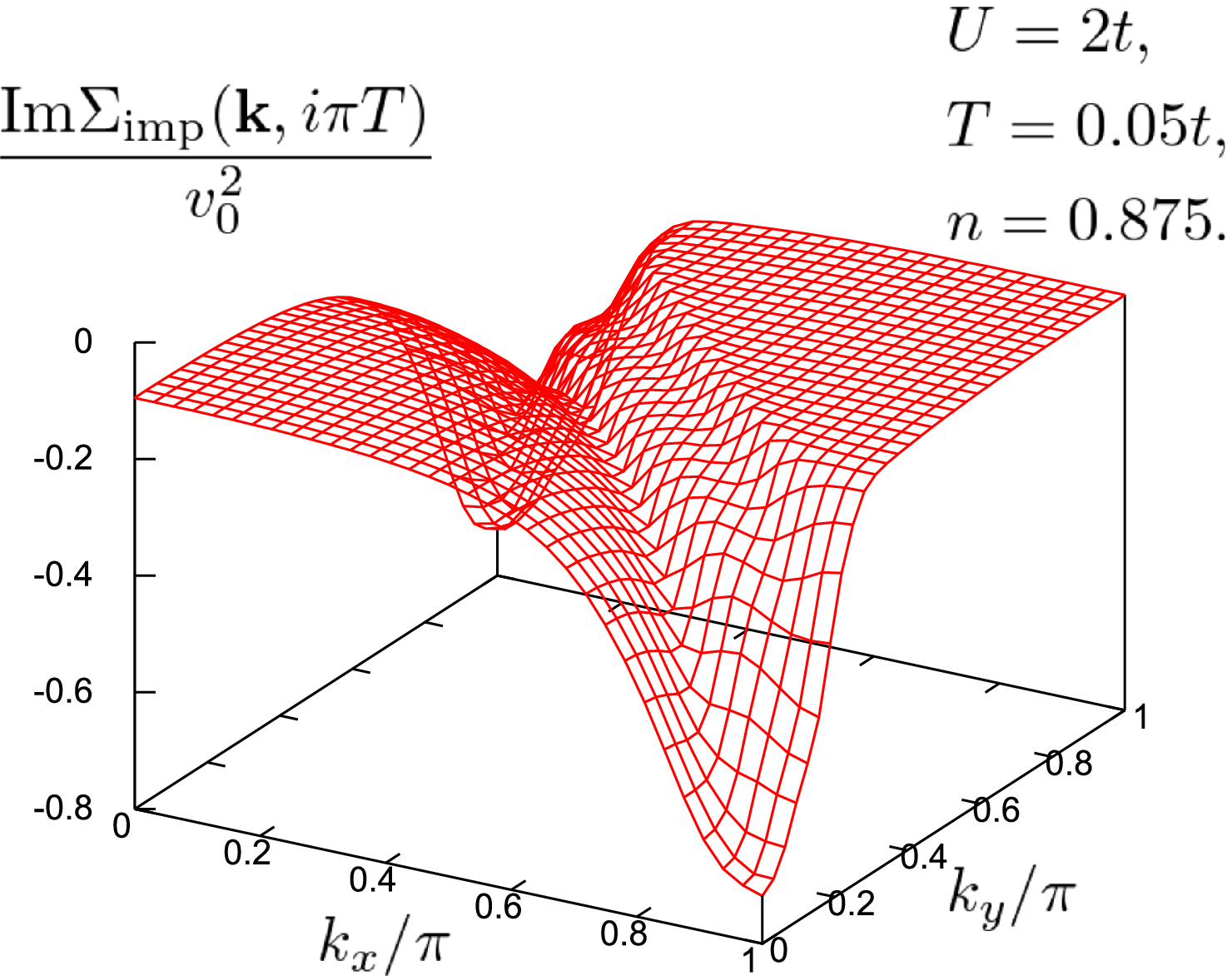}
 \end{center}
 \caption{Real and imaginary parts of the enhancement factor of the  Born self-energy 
$\Sigma_{\rm imp}({\bf k},i\pi T)/v_0^2$, with $\Sigma_{\rm imp}$ given by Eq.\ (\ref{eq:sg-Born2}), 
in the limit $v_0\to 0$.}
\label{fig-sg-imp}
\end{figure}
%%%%%%%%%%%%%%%%%%%%%%%%%%%%%%%%%%%%%%%%%%%%%%%%%%%%%%%%%%%%%%%%%%%%%%%%%%%%%%%
Figure \ref{fig-sg-imp} shows the real and imaginary parts of the Born self-energy, 
eq.\ (\ref{eq:sg-Born2}), 
in the first-order correction with respect to $v_{0}^{2}$.  On the other hand, 
the Born self-energy without vertex correction at $n=0.875$ becomes almost independent of 
${\bf k}$, in the low Matsubara frequency limit as
\begin{eqnarray}
{\rm Im}\Sigma_{\rm imp}({\bf k},i\epsilon_n)&\simeq&
-c_{\rm imp}v_0^2N_{\rm F}{\rm sign}(\epsilon_n)\nonumber\\
&=&-0.027\times v_0^2{\rm sign}(\epsilon_n).
\end{eqnarray}
This has essentially no ${\bf k}$ dependence.  One can see that the scattering amplitude 
Im$\Sigma_{\rm imp}({\bf k},i\pi T)$, is enhanced around the Fermi surface particularly 
in the ($0,\pi$) antinode direction.  Not only the imaginary part but also the real part 
of Im$\Sigma_{\rm imp}({\bf k},i\pi T)$ exhibits a visible ${\bf k}$ dependence.  
Since $c_{\rm imp}v_0^2$ is definitely positive, real-part correction works by reducing 
the next-nearest-neighbor hopping $t'$.
The above results display the enhancement of the scattering rate due to critical AF fluctuations.

Let us see the characteristic of the resulting doping phase diagram obtained by taking into account 
the impurity potential renormalized by the vertex correction.  
%%%%%%%%%%%%%%%%%%%%%%%%%%%%%%%%%%%%%%%%%%%%%%%%%%%%%%%%%%%%%%%%%%%%%%%%%%%%%%%
\begin{figure}
 \begin{center}
  \includegraphics[scale=0.5]{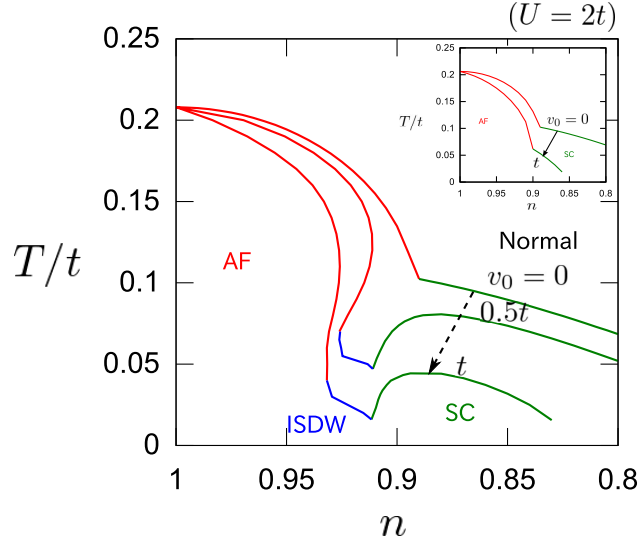}
 \end{center}
 \caption{(Color) Doping phase diagram of disordered extended Hubbard model, eq. (\ref{eq:exhubbard}), 
with vertex correction for impurity potential. $v_0$ denotes the strength of the bare impurity 
potential.  Each line represents the transition temperature of the 
antiferromagnetism (AF) in red, the incommensurate spin density wave (ISDW) in blue, and the 
superconductivity (SC) in green. Inset: phase diagram without vertex correction for 
impurity potential.}
\label{fig-pd-imp}
\end{figure}
%%%%%%%%%%%%%%%%%%%%%%%%%%%%%%%%%%%%%%%%%%%%%%%%%%%%%%%%%%%%%%%%%%%%%%%%%%%%%%%
Figure \ref{fig-pd-imp} shows the resulting doping phase diagram in the extended Hubbard model 
with disorder, eq. (\ref{eq:exhubbard}). As we expected from previous discussions, 
we find a remarkable suppression effect on the transition temperatures $T_{\rm N}$ and $T_{\rm c}$ 
around the competing region of antiferromagnetism and 
superconductivity.  This is due to the enhancement of the impurity potential renormalized by 
AL-type vertex correction. In fact, such a strong suppression of $T_{\rm N}$ and 
$T_{\rm c}$ cannot be understood without the vertex correction shown in the inset of 
Fig. \ref{fig-pd-imp}, for which no such suppression appears  

%1st order?
In the phase diagram in Fig. \ref{fig-pd-imp}, the AF phase shows a reentrant feature as the 
temperature is lowered. However, because the coefficient of the quartic term $\gamma_4$ in 
the Ginzburg-Landau free energy takes a negative value in the low-temperature region 
$T<0.524|\mu|$, the AF transition is expected to be of the first order in the low-temperature 
region where $dT_{\rm N}(n)/dn<0$.
However, in order to study the reality of the reentrant feature or the
first-order transition, a more reliable theory such as the $t$-matrix theory
or a theory beyond mean-field analysis is required.
For an incommensurate spin density wave transition, it is difficult to
conclude the order of the transition because it requires more detailed
analysis of free-energy structures considering couplings among multiple
ordering vectors.
%%%%%%%%%%%%%%%%%%%%%%%%%%%%%%%%%%%%%%%%%%%%%%%%%%%%%%%%%%%%%%%%%%%%%%%%%%%%%%%
\begin{figure}
 \begin{center}
  \includegraphics[scale=0.5]{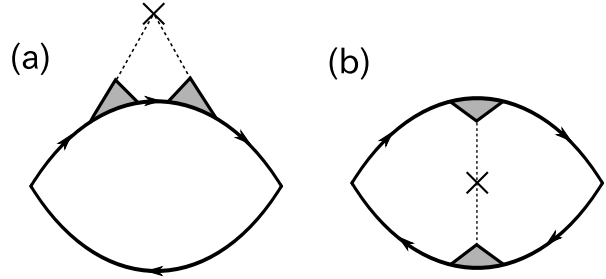}
 \end{center}
 \caption{(a) Self-energy correction and (b) vertex correction for the spin susceptibility by 
the renormalized impurity potential. In this study, we ignore the latter vertex diagram (b).}
\label{fig-vertex-ph}
\end{figure}
%%%%%%%%%%%%%%%%%%%%%%%%%%%%%%%%%%%%%%%%%%%%%%%%%%%%%%%%%%%%%%%%%%%%%%%%%%%%%%%
%change in AF
The $v_0$ dependence of the doping phase diagram in Fig. \ref{fig-pd-imp} reproduces the 
tendency that both antiferromagnetism and superconductivity disappear around the competing 
region of the AF and SC phases 
when we move from five-layered compounds \cite{Mukuda2006a,Mukuda2006b,Mukuda2008,Mukuda2009} to the  
double-layered compound YBCO \cite{Haug2010}. Moreover, the resulting phase diagram reproduces 
the emergence of the incommensurate SDW phase observed in YBCO.~\cite{Haug2010}  
However, we must be careful with the application of our theory to more disordered cuprates 
such as LSCO.  
In the underdoped LSCO, which experiences a stronger disorder, the possibility of a spin-glass state 
\cite{Chou1995}, a charge-glass state \cite{Raicevic2008}, and the Anderson localization due to 
the two-dimensional character should be discussed. Because the Born approximation is not suitable for 
a strongly disordered system, our theory cannot describe such insulating characteristics 
originating from a strong inhomogeneity. Thus, our theory should be applied to systems ]
having a moderate (out-of-plane) disorder.

In contrast with the experimental phase diagram, the AF phase is more robust against the impurity 
potential than the SC phase in our results.  
The actual critical doping rate for the disappearance of antiferromagnetism is obtained as 
$\delta_{\rm c}\sim0.12$ for five-layered cuprates, $0.05$ for YBCO, and $0.02$ for LSCO.  
In our analysis, however, it is obtained as $\delta_{\rm c}\sim 0.07$ for $v_0=t$ where the SC 
transition temperature $T_{\rm c}$ is lowered below half of the $T_{\rm c}$ for $v_0=0$.
 There are two reasons for this discrepancy. 
First, we have ignored vertex corrections for the AF susceptibility as shown in 
Fig. \ref{fig-vertex-ph}(b).  
Different from that for the anisotropic superconductivity, the vertex correction for the AF 
susceptibility has a role of destabilizing the N\'{e}el order together with self-energy 
corrections, as shown in Fig. \ref{fig-vertex-ph}(a).  
However, we had to ignore this kind of vertex correction because of the long 
computational time required to evaluate the so-called diffuson diagram in which vertex 
corrections are summed up to the infinite ladder for the renormalized impurity potential.
Second, both the effects of strong spin fluctuations and strong pairing fluctuations are missing 
in our mean-field analysis. Note that cuprates have a larger Coulomb repulsion such as 
$U\sim 6t$ or $8t$. In this region, the pairing interaction goes to a relatively strong 
coupling regime.  In the presence of atrong pairing interaction, a Cooper pair with a 
short coherence length is realized. In such a case, the SC state becomes robust against 
impurities in general.~ \cite{Yanase2006}

Although electrons in cuprates experience a strong Coulomb repulsion, we have analyzed  
in the weak-coupling situation that $U=2t$ because the mean-field approximation gives a greatly 
overestimated $T_{\rm N}$ for large-$U$ regions.
Our weak-coupling assumption can be justified to some extent by regarding $U$ as an
effective low-energy vertex function in the spin channel. Even if the bare $U$ is as large 
as the bandwidth $8t$, the lo-energy effective vertex function is suppressed to $\sim2t$ 
at half-filling in the two-particle 
self-consistent theory\cite{Vilk1997} and self-consistent fluctuation theory \cite{Kusunose2010}. 
These theories are attempts to determine a constant $U_{\rm sp}$ which is an approximate spin-channel 
vertex function in a reliable way. The suppression of the effective interaction in the spin 
channel has also been observed in the pseudo-potential parquet 
approximations. \cite{Bickers1991} These saturating 
phenomena of magnetic interaction are well known as the Kanamori-Br\"uckner screening effect 
in the problem of the ferromagnetic transition in itinerant electron systems.
Owing to the suppression of the effective interaction triggering antiferromagnetism, our 
mean-field analysis is expected to give a qualitatively reasonable description for cuprates.

In addition, we have not considered the next-nearest neighbor hopping $t'$ and the 
third-nearest neighbor hopping $t''$, which are necessary for reproducing the 
quasi-particle dispersions in cuprates. In weak-coupling cases, however, the effect 
of $t'$ makes the AF phase appear only away from the 
half-filling. \cite{Lin1987} Then, we would have to require a reliable calculation method in the 
large-$U$ regime for the purpose of treating an actual band dispersion in cuprates.

\section{Conclusions}
We have investigated the doping phase diagram of high-$T_{\rm c}$ cuprate superconductors with a 
moderate out-of-plane disorder in order to understand the mechanism by which the coexistence of 
antiferromagnetism and superconductivity exists in five-layered cuprates, but is suppressed 
toward double- and single-layered cuprates as the number of layers decreases.  
We hypothesized, in solving this problem, that the impurity potential is enhanced by the many-body 
corrections in the underdoped region, as shown in Fig. \ref{fig-schematic}. In fact, the experiments 
on the charge susceptibility and discussion on the basis of the Fermi liquid theory suggest that 
the enhancement of the impurity potential should occur in the underdoped cuprates through 
the many-body effects leading to the enhancement of the charge susceptibility.

On the basis of the microscopic theory taking the scattering process through the 
renormalized impurity potential into 
account, we have shown that the effective impurity potential can actually be enhanced in the 
underdoped region by the AL-type process which also causes the enhancement of the uniform 
charge susceptibility. Although the AL-type corrections to the impurity potential gives 
a $1/z$ factor, they lead to the remarkable effects 
that the coexistence region of AF and SC decreases in size and 
that the peak of the SC transition temperature 
$T_{\rm c}$ moves away from the AF critical point with increasing the disorder strength in the doping 
phase diagram.

As can be seen in Fig. \ref{fig-pd-imp}, the enhancement of the disorder effect is actually 
substantiated within the analysis by the mean-field approximation and Born approximation.
The resulting doping phase diagram can account for the phase diagram of YBCO, but not for 
the emergence of the glassy state in LSCO, relatively strongly disordered cuprates. 
This problem stems from the limited available range 
of the Born approximation, however, it allows the validity of the concept of the renormalization 
of disorder potential due to strong electron correlations.

\section*{Acknowledgments}
This work is supported in part by a Grant-in-Aid for Specially Promoted Research (No. 20001004) 
and by a Grant-in-Aid for Scientific Research on Innovative Areas ``Heavy Electrons'' 
(No. 20102008) from the Ministry of Education, Culture, Sports, Science and Technology.
One of us (H.T.) is supported by the Japan Society for the Promotion of Science through 
a Research Fellowship for Young Scientists.

\appendix
%%%%%
\section{Cubic mode-mode coupling strength among charge fluctuations and AF fluctuations}\label{app:cubic}
%%%%%
The coupling strength among one charge fluctuation field at ${\bf q}=0$ and two AF fluctuation fields at 
${\bf Q}^*=(\pi,\pi)$ is given by
\begin{eqnarray}
 \gamma_3(0,{\bf Q}^*;0)=T\sum_{{\bf k},n}\left(\frac{1}{i\epsilon_n-\epsilon({\bf k})+\mu}\right)^2
\frac{1}{i\epsilon_n-\epsilon({\bf k+Q}^*)+\mu}.
\end{eqnarray}
Here, we suppose that the single-particle dispersion in the square lattice is 
given as $\epsilon_{\bf k}=-2t(\cos k_x+\cos k_y)$. This dispersion satisfies the perfect nesting 
condition $\epsilon_{\bf k+Q}=-\epsilon_{\bf k}$.  
Then, the coupling $\gamma_3$ is rewritten as
\begin{eqnarray}
 \gamma_3&\sim&T\sum_{n}N_{\rm F}\int_{-\infty}^{\infty}d\epsilon
\left(\frac{1}{i\epsilon_n-\epsilon+\mu}\right)^2
\frac{1}{i\epsilon_n+\epsilon+\mu}\nonumber\\
&=&\frac{i\pi N_{\rm F}T}{2}\sum_{n\ge 0}\left[\frac{1}{\left(i\epsilon_n-\mu\right)^2}-
\frac{1}{\left(i\epsilon_n+\mu\right)^2}\right]\nonumber\\
&=&-2\pi N_{\rm F}T\mu\sum_{n\ge 0}\frac{\epsilon_n}{\left(\epsilon_n^2+\mu^2\right)^2}\,.
\label{eq:A2}
\end{eqnarray}
In the region $|\mu|\ll\pi T$, $\gamma_3$ can be expanded by $\mu$. The leading term is obtained as
\begin{eqnarray}
 \gamma_3&\sim&-\frac{2N_{\rm F}\mu}{\pi^2T^2}\sum_{n\geq 0}\frac{1}{(2n+1)^3}\nonumber\\
&=&-\frac{7\zeta(3)N_{\rm F}\mu}{4\pi^2T^2},
\end{eqnarray}
which is linear in $\mu$.  
In contrast, in the region $|\mu|\gg\pi T$, the summation can be approximated by an integral 
with respect to $\epsilon$ as
\begin{eqnarray}
 \gamma_3&\sim& -N_{\rm F}\mu\int_{0}^{\infty}d\epsilon
\frac{\epsilon}{\left(\epsilon^2+\mu^2\right)^2}\nonumber\\
&=&-\frac{N_{\rm F}}{2\mu}.
\end{eqnarray}
The right-hand side of eq.\ (\ref{eq:A2}) is equal to 
$(N_{\rm F}/4\pi T){\rm Im}\psi^{\prime}(1/2+i\mu/2\pi T)$, where $\psi^{\prime}$ is the 
tri-gamma function.  
Therefore, $|\gamma_3|$ reaches its maximum at $\mu\sim\mu_{\rm max}$, which corresponds to 
the condition 
${\rm Re}\psi^{(2)}(1/2+i\mu_{\rm max}/2\pi T)=0$, $\psi^{(2)}$ being the tetra-gamma function, 
because the following identity holds
\begin{equation}
{\partial\over \partial\mu}{\rm Im}\psi^{\prime}\left({1\over 2}+{i\mu\over2\pi T}\right)
={1\over2\pi T}{\rm Re}\psi^{(2)}\left({1\over 2}+{i\mu\over2\pi T}\right).
\end{equation}
Then, $|\mu_{\rm max}|$ is given by $|\mu_{\rm max}|\simeq 1.91 T$, (See Appendix C).

\section{Numerical procedure for calculating Aslamazov-Larkin term}
The FFT algorithm is useful in the evaluation of the AL term where the convolution 
integrals are included. 
In this appendix, we explain how it is applied to the calculation of AL terms. We consider 
the following AL term in the renormalization factor for the impurity potential:
\begin{eqnarray}
 \Lambda_{\rm AL}({\bf k},i\epsilon_n;{\bf q})
&=&T^2\sum_{{\bf k}',i\epsilon_n',{\bf q}',i\omega_m}G({\bf k}',i\epsilon_n')
G({\bf k}'+{\bf q},i\epsilon_n')
G({\bf k}'+{\bf q}',i\epsilon_n'+i\omega_m)\nonumber\\
& &\times\Gamma_{\rm s}({\bf q'},i\omega_m)\Gamma_{\rm s}({\bf q}-{\bf q'},i\omega_m)
G({\bf k}+{\bf q}',i\epsilon_n+i\omega_m).\label{eq:ap-AL}
\end{eqnarray}
It is more convenient to directly calculate the three-point vertex function $\lambda_{\rm AL}$ than 
the four-point vertex $\Gamma_{\rm 1,AL}$ because it is expressed by fewer momentum indices.  
The integral in the AL term can be rapidly performed with the use of the FFT at a fixed ${\bf q}$.
For this purpose, we disassemble eq. (\ref{eq:ap-AL}) as 
\begin{eqnarray}
 A({\bf k},i\epsilon;{\bf q})&=&
G({\bf k},i\epsilon_n)G({\bf k}+{\bf q},i\epsilon_n),\label{eq:ap-AL2}\\
B({\bf q}',i\omega_m;{\bf q})&=&
\sum_{{\bf k},i\epsilon_n}A({\bf k},i\epsilon_n;{\bf q})
G({\bf k}+{\bf q}',i\epsilon_n+i\omega_m),\label{eq:ap-AL3}\\
C({\bf q}',i\omega_m;{\bf q})&=&
B({\bf q}',i\omega_m;{\bf q})\Gamma({\bf q'},i\omega_m)\Gamma({\bf q}-{\bf q'},i\omega_m),\\
 \Lambda_{\rm AL}({\bf k},i\epsilon_n,{\bf q})&=&
\sum_{{\bf q}',i\omega_m}C({\bf q}',i\omega_m;{\bf q})
G({\bf k}+{\bf q}',i\epsilon_n+i\omega_m).\label{eq:ap-AL5}
\end{eqnarray}
We can obtain the renormalization factor $\Lambda_{\rm AL}$ through 
eqs. (\ref{eq:ap-AL2}) - (\ref{eq:ap-AL5}).
Note that eqs. (\ref{eq:ap-AL3}) and (\ref{eq:ap-AL5}) are expressed as the convolution form.  
Thus, its numerical calculation requires only the number of times of ${\cal O}(N\log N)$ for 
a fixed transfer momentum ${\bf q}$.
Moreover, we can obtain the AL-type correction for the static charge susceptibility from the 
three-point function $\Lambda_{\rm AL}$ as
\begin{eqnarray}
 \Delta\chi_{\rm c,AL}({\bf q})=
\sum_{{\bf k},i\epsilon_n}\Lambda_{\rm AL}({\bf k},i\epsilon_n;{\bf q})
G({\bf k},i\epsilon_n)G({\bf k}+{\bf q},i\epsilon_n).
\end{eqnarray}

%%%%%
\section{Quartic AF mode-mode coupling strength}\label{app:quartic}
%%%%%
In this appendix, the quartic coupling strength among the four AF spin fluctuation modes is 
calculated.  This coupling strength is equivalent to the fourth-order coefficient of the 
Ginzburg-Landau free energy for an AF order parameter.

We assume the same single-particle dispersion used in Appendix \ref{app:cubic}. 
The quartic coupling $\gamma_{4}$ of AF spin fluctuation modes at ${\bf Q}^*=(\pi,\pi)$  is evaluated as
\begin{eqnarray}
\gamma_{4}&=&T\sum_{{\bf k},n}\frac{1}{(i\epsilon_n-\epsilon({\bf k})+\mu)^2
(i\epsilon_n-\epsilon({\bf k}+{\bf Q}^*)+\mu)^2}\nonumber\\
&=&T\sum_{{\bf k},n}\frac{1}{(i\epsilon_n-\epsilon({\bf k})+\mu)^2
(i\epsilon_n+\epsilon({\bf k})+\mu)^2}\nonumber\\
&\sim&T\sum_{n}N_{\rm F}\int^\infty_{-\infty} d\epsilon 
\frac{1}{(i\epsilon_n-\epsilon+\mu)^2(i\epsilon_n+\epsilon+\mu)^2}\nonumber\\
&=&-\frac{i\pi N_{\rm F}}{2}T\sum_{n\ge 0}\left[\frac{1}{(i\epsilon_n+\mu)^3}+
\frac{1}{(i\epsilon_n-\mu)^3}\right]
\nonumber\\
&=&-\frac{N_{\rm F}}{(4\pi T)^{2}}{\rm Re}\psi^{(2)}
\left({1\over 2}+{i|\mu|\over 2\pi T}\right),
\end{eqnarray}
where $\psi^{(2)}$ is the tetra-gamma function.  ${\rm Re}\psi^{(2)}(\frac{1}{2}+ix)$ changes 
its sign at $x\simeq 0.304$; i.e., ${\rm Re}\psi^{(2)}(\frac{1}{2}+ix)<0$\ ($>0$) for 
$x<0.304$\ ($x>0.304$).~\cite{MakiTsuneto}  
Thus, $\gamma_{4}$ changes its sign at $|\mu|\simeq 1.91T$ from positive to negative 
as $|\mu|$ increases; i.e., $\gamma_{4}$ is expected to be negative in the low-temperature region 
$T<0.524|\mu|$ in the underdoped region of the simple Hubbard model with $t'=0$.  
The negativity of the quartic coupling constant $\gamma_{4}$ 
leads to a first-order phase transition and makes higher-order AL terms positive.

\vskip24pt
\newpage
\noindent
{\bf Note added in proof:}

After the present paper had been accepted for publication, we realized that we 
had missed to refer previous works which reported an experimental aspects 
related to the physics discussed in the present paper.  

Effects of defects in correlated superconductors are reviewed in 
H.Alloul, J. Bobroff, M. Gabay, and P. Hirschfeld: Rev. Mod. Phys. 
{\bf 81} (2009) 45, in which the role of disorder in single-layered curate Hg1201 
compound is discussed.  The importance of taking into account the disorder 
for understanding the phase diagram, especially apparent competition between 
magnetism and superconductivity, of double-layered cuprate YBCO systems was 
discussed from experimental side in F. Rullier-Albenque, H. Alloul, F. Balakirev, and 
C. Proust: Europhys. Lett. {\bf 81} (2008) 37008. 

We thank Prof. Henri Alloul for reminding us of those works.  

\begin{thebibliography}{99}
%Phase diagram of LSCO
\bibitem{Keimer1992}
B. Keimer, N. Belk, R. J. Birgeneau, A. Cassanho, C. Y. Chen, M. Greven, M. A. Kastner, 
A. Aharony, Y. Endoh, 
R. W. Erwin, and G. Shirane: Phys. Rev. B {\bf 46} (1992) 14034.
%Hg multi-layered cuprate
\bibitem{Mukuda2006a}
H. Mukuda, M. Abe, Y. Araki, Y. Kitaoka, K. Tokiwa, T. Watanabe, A. Iyo, H. Kito, and Y. Tanaka: 
Phys. Rev. Lett. {\bf 96} (2006) 087001.
\bibitem{Mukuda2006b}
H. Mukuda, M. Abe, S. Shimizu, Y. Kitaoka, A. Iyo, Y. Kodama, H. Kito, Y. Tanaka, K. Tokiwa, and 
T. Watanabe: J. Phys. Soc. Jpn. {\bf 75} (2006) 123702.
\bibitem{Mukuda2008}
H. Mukuda, Y. Yamaguchi, S. Shimizu, Y. Kitaoka, P. Shirage, and A. Iyo: 
J. Phys. Soc. Jpn. {\bf 77} (2008) 124706.
\bibitem{Mukuda2009}
H. Mukuda, Y. Yamaguchi, S. Shimizu, Y. Kitaoka, P. Shirage, and A. Iyo: 
J. Phys. Conf. Ser. {\bf 150} (2009) 052176.
%coexistence in Hubbard model
\bibitem{Reiss2007}
J. Reiss, D. Rohe, and W. Metzner: Phys. Rev. B {\bf 75} (2007) 075110.
\bibitem{Jarrell2001}
M. Jarrell, T. Maier, M. Hettler, and A. Tahvildarzadeh: Europhys. Lett. {\bf 56} (2001) 563.
\bibitem{Aichhorn2006}
M. Aichhorn, E. Arrigoni, M. Potthoff, and W. Hanke: Phys. Rev. B {\bf 74} (2006) 024508.
\bibitem{Aichhorn2007}
M. Aichhorn, E. Arrigoni, M. Potthoff, and W. Hanke: Phys. Rev. B {\bf 76} (2007) 224509.
\bibitem{Senechal2005}
D. S\'{e}n\'{e}chal, P.-L. Lavertu, M.-A. Marois, and A.-M. S. Tremblay: Phys.
  Rev. Lett. {\bf 94} (2005) 156404.
\bibitem{Kobayashi2009}
K. Kobayashi, T. Watanabe, and H. Yokoyama: Physica C {\bf 470} (2010) S947.
\bibitem{Lichtenstein2000}
A. I. Lichtenstein and M. I. Katsnelson: Phys. Rev. B {\bf 62} (2000) R9283.
\bibitem{Capone2006}
M. Capone and G. Kotliar: Phys. Rev. B {\bf 74} (2006) 054513.
\bibitem{Kancharla2008}
S. S. Kancharla, B. Kyung, D. S\'{e}n\'{e}chal, M. Civelli, M. Capone, G. Kotliar, and 
A. -M. S. Tremblay: 
Phys. Rev. B {\bf 77} (2008) 184516.
%coexistence in t-J model
\bibitem{Inaba1996}
M. Inaba, H. Matsukawa, M. Saitoh, and H. Fukuyama: Physica C {\bf 257} (1996) 299.
\bibitem{Yamase2004}
H. Yamase and H. Kohno: Phys. Rev. B {\bf 69} (2004) 104526.
\bibitem{Himeda1999}
A. Himeda and M. Ogata: Phys. Rev. B {\bf 60} (1999) R9935.
\bibitem{Pathak2009a}
S. Pathak, V. B. Shenoy, M. Randeria, and N. Trivedi: Phys. Rev. Lett. {\bf 102} (2009) 027002.
%first principle in LSCO
\bibitem{Anisimov1992}
V. I. Anisimov, M. A. Korotin, J. Zaanen, and O. K. Anderson: Phys. Rev. Lett. {\bf 68} (1992) 345.
%in-plane disorder
\bibitem{Harashina1993}
H. Harashina, T. Nishikawa, T. Kiyokura, S. Shamoto, M. Sato, and K. Kakurai: 
Physica C {\bf 212} (1993) 142.
%out-of-plane disorder
\bibitem{Fujita2005}
K. Fujita, T. Noda, K. M. Kojima, H. Eisaki, and S. Uchida: Phys. Rev. Lett. {\bf 95} (2005) 097006.
\bibitem{Sugimoto2006}
A. Sugimoto, S. Kashiwaya, H. Eisaki, H. Kashiwaya, H. Tsuchiura, Y. Tanaka, K. Fujita, 
and S. Uchida: Phys. Rev. B {\bf 74} (2006) 094503.
\bibitem{Okada2008}
Y. Okada, T. Takeuchi, T. Baba, S. Shin, and H. Ikuta: J. Phys. Soc. Jpn. {\bf 77} (2008) 074714.
\bibitem{Hobou2009}
H. Hobou, S. Ishida, K. Fujita, M. Ishikado, K. M. Kojima, H. Eisaki, and S. Uchida: 
Phys. Rev. B {\bf 79} (2009) 064507.
%coexistence in YBCO
\bibitem{Haug2010}
D. Haug, V. Hinkov, Y. Sidis, P. Bourges, N. B. Christensen, A. Ivanov, T. Keller, C. T. Lin, 
and B. Keimer: 
New J. Phys. {\bf 12} (2010) 105006.
%AG formula
\bibitem{Abrikosov1961}
A. A. Abrikosov and L. P. Gor'kov: Sov. Phys. JETP {\bf 12} (1960) 1243.
%Fermi liquid theory on impurity scattering
\bibitem{Betbeder-Matibet1966} O. Betbeder-Matibet and P. Nozi\`eres: 
Ann. Phys. (N. Y.) {\bf 37} (1966) 17.
\bibitem{Landau1956} L. D. Landau: Sov. Phys. JETP {\bf 3} (1956) 920.
\bibitem{Miyake2001}
K. Miyake and H. Maebashi: J. Phys. Chem. Solids {\bf 62} (2001) 53.
\bibitem{Miyake2002a}
K. Miyake and O. Narikiyo: J. Phys. Soc. Jpn. {\bf 71} (2002) 867.
\bibitem{Miyake2002b}
 K. Miyake and H. Maebashi: J. Phys. Soc. Jpn. {\bf 71} (2002) 1007.
\bibitem{Maebashi2002}
H. Maebashi, K. Miyake, and C. Varma: Phys. Rev. Lett. {\bf 88} (2002) 226403.
%Monte Carlo - Uniform charge susceptibility
\bibitem{Otuka1990} H. Otsuka: J. Phys. Soc. Jpn. {\bf 59} (1990) 2916.
\bibitem{Furukawa1991} N. Furukawa and M. Imada: J. Phys. Soc. Jpn. {\bf 60} (1991) 3604.
\bibitem{Furukawa1992} N. Furukawa and M. Imada: J. Phys. Soc. Jpn. {\bf 61} (1992) 3331.
\bibitem{Furukawa1993} N. Furukawa and M. Imada: J. Phys. Soc. Jpn. {\bf 62} (1993) 2557.
\bibitem{Watanabe2004a}
S. Watanabe and M. Imada: J. Phys. Soc. Jpn. {\bf 73} (2004) 1251.
%Photoemission - Uniform charge susceptibility
\bibitem{Ino1997}
A. Ino, T. Mizokawa, A. Fujimori, K. Tamasaku, H. Eisaki, S. Uchida, T. Kimura, T. Sasagawa, 
and K. Kishio: Phys. Rev. Lett. {\bf 79} (1997) 2101.
\bibitem{Harima2001}
N. Harima, J. Matsuno, A. Fujimori, Y. Onose, Y. Taguchi, and Y. Tokura: 
Phys. Rev. B {\bf 64} (2001) 220507.
\bibitem{Harima2003}
N. Harima, A. Fujimori, T. Sugaya, and I. Terasaki: Phys. Rev. B {\bf 67} (2003) 172501.
\bibitem{Yagi2006}
H. Yagi, T. Yoshida, A. Fujimori, Y. Kohsaka, M. Misawa, T. Sasagawa, H. Takagi, M. Azuma, 
and M. Takano: Phys. Rev. B {\bf 73} (2006) 172503.
\bibitem{Ikeda2010}
M. Ikeda, M. Takizawa, T. Yoshida, A. Fujimori, K. Segawa, and Y. Ando: 
Phys. Rev. B {\bf 82} (2010) 020503(R).
%DMFT 1d hubbard
\bibitem{Capone2004}
M. Capone, M. Civelli, S. Kancharla, C. Castellani, and G. Kotliar: 
Phys. Rev. B {\bf 69} (2004) 195105.
%Aslamazov - Larkin term on charge susceptibility
\bibitem{Miyake1994} K. Miyake and O. Narikiyo: J. Phys. Soc. Jpn. {\bf 63} (1994) 2042. 
\bibitem{Aslamazov1968}
L. G. Aslamazov and A. I. Larkin: Sov. Phys. Solid State {\bf 10} (1968) 875.
%conserving approximation
\bibitem{Baym1961} G. Baym and L. Kadanoff: Phys. Rev. {\bf 124} (1961) 287.
\bibitem{Bickers1989} N. E. Bickers, D. J. Scalapino, and S. R. White: Phys. Rev. Lett. 62 (1989) 961.
\bibitem{AGD1965}A. A. Abrikosov, L. P. Gor'kov, and I. Ye. Dzyaloshinskii: 
{\it Quantum Field Theoretical Methods in Statistical Physics} (Pergamon Press, Oxford, 1965) 
2nd ed., Sec. 19.6. 
%spin-glass,charge-glass
\bibitem{Chou1995}
F. C. Chou, N. R. Belk, M. A. Kastner, R. J. Birgeneau, and A. Aharony: 
Phys. Rev. Lett. {\bf 75} (1995) 2204.
\bibitem{Raicevic2008}
I. Rai\u{c}evi\'{c}, J. Jaroszy\'{n}ski, D. Popovi\'{c}, C. Panagopoulos, and T. Sasagawa: 
Phys. Rev. Lett. {\bf 101} (2008) 177004.
\bibitem{Yanase2006}
Y. Yanase: J. Phys. Soc. Jpn. {\bf 75} (2006) 124715.
%effective U
\bibitem{Vilk1997}
Y. M. Vilk and A.-M. S. Tremblay: J. Phys. I (France) {\bf 7} (1997) 1309.
\bibitem{Kusunose2010}
H. Kusunose: J. Phys. Soc. Jpn. {\bf 79} (2010) 094707.
\bibitem{Bickers1991}
N. Bickers and S. White: Phys. Rev. B {\bf 43} (1991) 8044.
\bibitem{Lin1987} H. Q. Lin and J. E. Hirsch: Phys. Rev. B {\bf 35} (1987) 3359.
\bibitem{MakiTsuneto}
K. Maki and T. Tsuneto: Prog. Theor. Phys. {\bf 31} (1964) 945. 
\end{thebibliography}
\end{document}